\documentclass[pra,aps,twocolumn,superscriptaddress,showpacs,eqsecnum,nofootinbib]{revtex4}

\usepackage{amsmath}
\usepackage{graphicx}
\usepackage{color}
\usepackage{hyperref}
\usepackage{ulem}

\begin{document}

\title{Dynamics of a spin qubit in an optical dipole trap}

\author{L.V. Gerasimov}
\email{gerasimov\textunderscore lv@spbstu.ru}
\affiliation{Faculty of Physics, M.V. Lomonosov Moscow State University, Leninskiye Gory 1-2, 119991, Moscow, Russia}
\affiliation{Quantum Technologies Center, M.V. Lomonosov Moscow State University, Leninskiye Gory 1-35, 119991, Moscow, Russia}
\affiliation{Center for Advanced Studies, Peter the Great St. Petersburg Polytechnic University, 195251, St. Petersburg, Russia}
\author{R.R. Yusupov}
\affiliation{Faculty of Physics, M.V. Lomonosov Moscow State University, Leninskiye Gory 1-2, 119991, Moscow, Russia}
\affiliation{Quantum Technologies Center, M.V. Lomonosov Moscow State University, Leninskiye Gory 1-35, 119991, Moscow, Russia}
\author{I.B. Bobrov}
\affiliation{Faculty of Physics, M.V. Lomonosov Moscow State University, Leninskiye Gory 1-2, 119991, Moscow, Russia}
\affiliation{Quantum Technologies Center, M.V. Lomonosov Moscow State University, Leninskiye Gory 1-35, 119991, Moscow, Russia}
\author{D. Shchepanovich}
\affiliation{Faculty of Physics, M.V. Lomonosov Moscow State University, Leninskiye Gory 1-2, 119991, Moscow, Russia}
\affiliation{Quantum Technologies Center, M.V. Lomonosov Moscow State University, Leninskiye Gory 1-35, 119991, Moscow, Russia}
\author{E.V. Kovlakov}
\affiliation{Faculty of Physics, M.V. Lomonosov Moscow State University, Leninskiye Gory 1-2, 119991, Moscow, Russia}
\affiliation{Quantum Technologies Center, M.V. Lomonosov Moscow State University, Leninskiye Gory 1-35, 119991, Moscow, Russia}
\author{S.S. Straupe}
\email{straups@quantum.msu.ru}
\affiliation{Faculty of Physics, M.V. Lomonosov Moscow State University, Leninskiye Gory 1-2, 119991, Moscow, Russia}
\affiliation{Quantum Technologies Center, M.V. Lomonosov Moscow State University, Leninskiye Gory 1-35, 119991, Moscow, Russia}
\author{S.P. Kulik}
\affiliation{Faculty of Physics, M.V. Lomonosov Moscow State University, Leninskiye Gory 1-2, 119991, Moscow, Russia}
\affiliation{Quantum Technologies Center, M.V. Lomonosov Moscow State University, Leninskiye Gory 1-35, 119991, Moscow, Russia}
\author{D.V. Kupriyanov}
\affiliation{Quantum Technologies Center, M.V. Lomonosov Moscow State University, Leninskiye Gory 1-35, 119991, Moscow, Russia}
\affiliation{Center for Advanced Studies, Peter the Great St. Petersburg Polytechnic University, 195251, St. Petersburg, Russia}

\begin{abstract}
\noindent We present a theoretical investigation of coherent dynamics of a spin qubit encoded in hyperfine sublevels of an alkali-metal atom in a far off-resonant optical dipole trap. The qubit is prepared in the ``clock transition'' utilizing the Zeeman states with zero projection of the spin angular momentum. We focus on various dephasing processes such as the residual motion of the atom, fluctuations of the trapping field and its incoherent scattering and their effects on the qubit dynamics. We implement the most general fully-quantum treatment of the atomic motion, so our results remain valid in the limit of close-to-ground-state cooling with low number of vibrational excitations. We support our results by comparison with an experiment showing reasonable correspondence with no fitting parameters. 


\end{abstract}


\maketitle

\section{Introduction}\label{Section_I}

Recent progress in laser cooled atoms, confined in far off-resonance optical dipole potentials, gives rise to development of an excellent experimental platform for quantum computing \cite{Saffman2019,Lukin2019} and quantum simulation \cite{Browaeys_Nature2016,Lukin_Nature2019,Browaeys_NaturePhys2020}. A possible quantum machine based on optically trapped neutral atoms relies on coherent manipulation of qubits, in which information is encoded in Zeeman-insensitive clock states. The coherence time of the neutral atom hyperfine qubits is a crucial parameter to judge the usefulness of the system for achieving high fidelity in single- and two-qubit gates. Moreover, long coherence times are of great importance for scaling the number of atomic qubits without limiting gate fidelities. 

The atomic hyperfine coherence would never degrade in an ideal scenario of a closed quantum system cooled down to its ground vibrational state. However, such conditions can never be perfectly achieved. A spin qubit based on an alkali-metal atom trapped in an optical dipole trap is thus an open quantum system, characterized by its residual motion and the coupling to the environment \cite{Zurek1982}. This coupling is responsible for atomic heating and decoherence, which results in the decay of a pure superposition state to a statistical mixture. Fundamental sources of heating by the trapping light are the spontaneous scattering of trap photons \cite{Heinzen1994,Grimm2000} as well as laser intensity noise and beam-pointing fluctuations \cite{LaserNoise1997}, which can also cause dephasing of hyperfine coherences of trapped atoms. Currently many important aspects of spin qubit decoherence are understood, explained and discussed in a number of experimental and theoretical works \cite{Kimble2000,Meschede2005,Polzik2008}. Convenient methods for coherent addressing and manipulating single atoms in optical tweezer and neutral atoms stored in optical lattices with single site resolution were developed in recent experiments \cite{Bloch2006,Grangier2007,Meshede2010}. In a number of experimental works high-fidelity single-qubit gates based on neutral atoms trapped in 2D and 3D arrays were performed and characterized \cite{Saffman2015,Weiss2016,Mingsheng2018}. However, even higher fidelities and better control of individual atomic qubits states are required for effective implementation of two-qubit entangling gates, which still poses a challenge in these systems.

In this paper we analyze dynamics of an atomic qubit localized in an optical dipole trap. The case of an atom moving in a tweezer potential raises a question of how the trapping field affects the qubit dynamics due to the intensity fluctuations and incoherent scattering effect. We assume that in a typical experimental scenario the microtrap is originally loaded from an atomic ensemble, prepared in a magneto-optical trap, and after a stage of molasses cooling still has a relatively high temperature and, as a consequence, a high mean vibrational number $\bar{v} > 1$ for each mode. This residual thermal motion results in Doppler broadening of the Rydberg excitation lasers and therefore it limits the fidelity of two-qubit Rydberg gates which ideally require ground-state cooling for effective operation. It also affects single-qubit gates due to differential light-shift of the qubit levels \cite{Derevianko_PRA2010}. 

Recent works show impressive experimental progress in ground-state cooling of a single atom by the Raman sideband cooling (RSC) technique \cite{Regal_PRX2012,Lukin2013,Andersen2017}. A set of critical requirements for the parameters of the RSC protocol was suggested and the Raman passage was simulated numerically in a three-dimensional configuration in \cite{RSC2019}. RSC allows one to reach the regime of low vibrational excitation, where the atomic motion has to be described quantum mechanically. At the same time, most of the models of motional decoherence in a microscopic dipole trap are based on a classical treatment of the atomic motion \cite{Meschede2005,Weinfurter_PRA2011}. This motivates us to develop a numerical approach to describe spin qubit dynamics with a fully quantum treatment of the atomic motion. Such model would be applicable when the trap oscillator is weakly excited $\bar{v} \sim 1$, which is of the most practical interest for implementation of high fidelity multiqubit gates.
 
The paper is organized as follows. In Section \ref{Section_II} we perform a general description of our numerical treatment of coherence dynamics of an atomic qubit confined with a dipole trap. In our theoretical approach we follow the parameters of our experimental setup. Details of the experiment are described in Section \ref{Section_III}. In Section \ref{Section_IV} we illustrate the results of our numerical simulations and compare them with the experiment. In Appendices \ref{Appendix_A} and \ref{Appendix_B} we introduce our formalism of coherent dynamics of the trapped atom and some dephasing mechanisms arising from its interactions with the environment in more details.

\section{Theoretical framework}\label{Section_II}

\subsection{Basic assumptions}
\begin{figure}[t]
\scalebox{0.4}{\includegraphics*{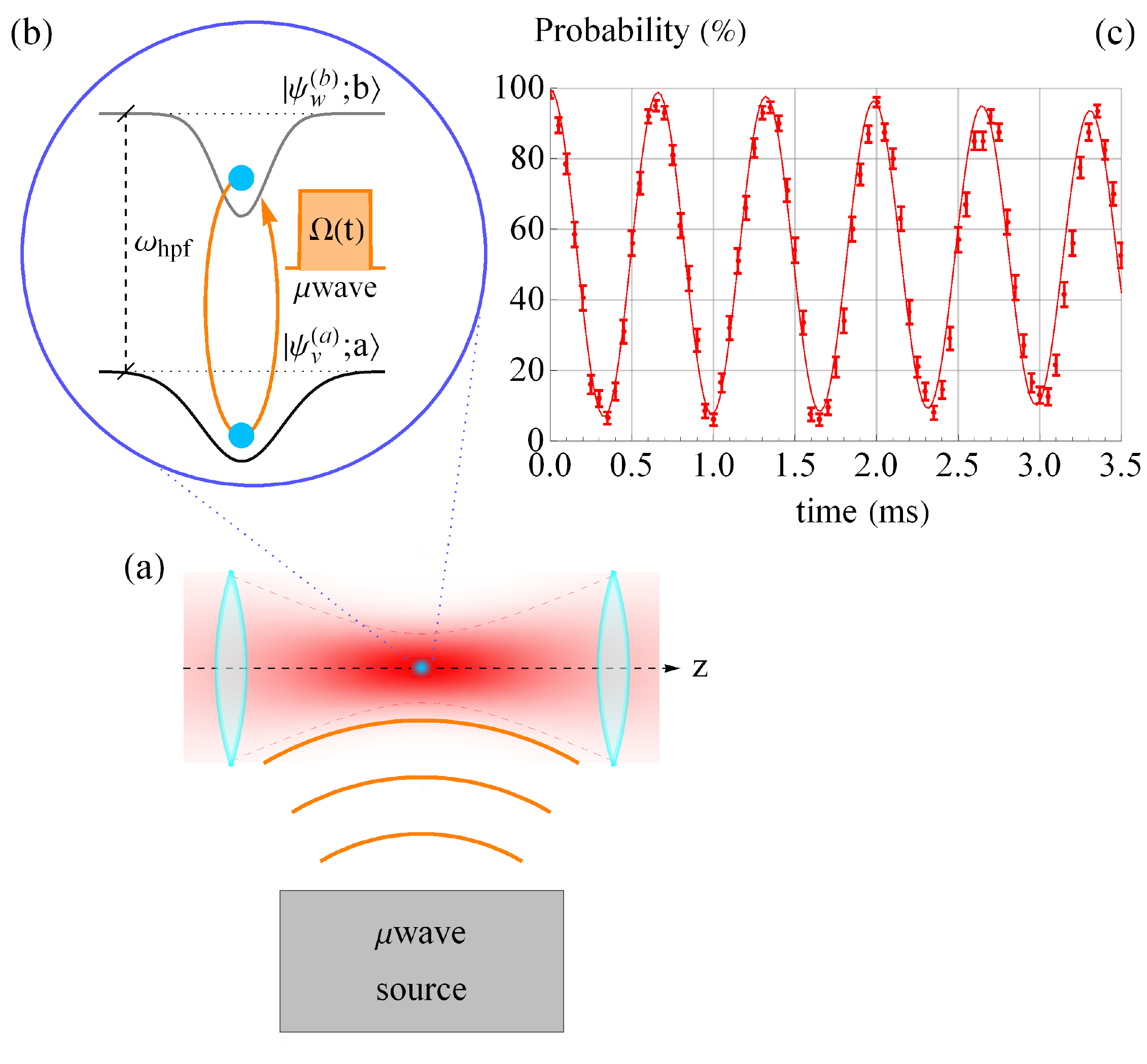}}
\caption{(Color online) Schematic of the qubit preparation: (a) The alkali-metal atom is loaded into the dipole trap and localized near the focal point of the laser beam. It is optically pumped onto one of the the clock states $|a\rangle = |F_{-}; 0\rangle$ and $|b\rangle = |F_{+},0\rangle$. The clock transition is driven by a microwave field having a Rabi frequency $\Omega(t)$ and linearly polarized along the quantization axis $z$; (b) The trapping potentials are slightly different for the spin states, which induces dephasing in the qubit dynamics; (c) The example of Rabi oscillations on the $|a\rangle \to |b\rangle$ transition recorded (points with error bars) and calculated (curve) for the beginning stage of the coherent dynamics driven by the stationary microwave field for the well depth $|U_0| \simeq 300\, \mu$K and for the temperature $T = 40\, \mu$K. }
\label{fig1}%
\end{figure}

\noindent Imagine a collection of alkali-metal atoms localized in microscopic optical dipole traps and configured as an ordered 2D lattice in space. In a typical setup such a lattice of tweezers can be prepared with separation about a few microns and with a typical atomic lifetime of a few seconds. Each of the atoms can be deterministically pumped onto the clock transition between the two Zeeman states $|F_{-},M_{0}=0\rangle\equiv|a\rangle$ and $|F_{+},M_{0}=0\rangle\equiv|b\rangle$, where $F_{\pm}=I\pm 1/2$ are the total spin angular momenta for the upper and lower hyperfine sublevels respectively; $I$ is the nuclear spin. These specific states have a zero value of the angular momentum projection to the quantization axis $M_0=0$ and no first-order Zeeman shifts of their energy levels. So the qubit levels are separated by approximately the same reference hyperfine splitting $\hbar\omega_{\mathrm{hpf}}$ (Fig. \ref{fig1} a,b) for all the atoms. 

The pair of states $|a\rangle$ and $|b\rangle$ are perfectly adjusted for the preparation of superposition states, which can be done by a short $\pi/2$ microwave pulse (in quantum information theory commonly referred as an Hadamard gate). In an ideal scenario, if any atom, isolated from the environment, is superposed between these spin states such a qubit system will never degrade. Indeed, the natural decay rate is given by the magnetic dipole transition probability \cite{BERESTETSKII19821}
\begin{equation}
\Gamma_0 = \frac{4 I g^2}{3(2I+1)}\frac{\omega^3_{\rm hpf}\mu^2_{\rm B}}{\hbar c^3} ,
\label{2.1}
\end{equation}
where $\mu_{\rm B}$ is the Bohr magneton, $g=-2$ is the electron $g$-factor, and $\omega_{\rm hpf} = \omega_{ba} = 2\pi\cdot6.83\ldots$GHz is the frequency of the ground state hyperfine splitting, taken for ${}^{87}$Rb in our example. This calculates to $\Gamma_0\sim 10^{-13}$ Hz, which is, in fact, an effectively zero rate for any duration of quantum processing. However this fantastically slow rate could never be approached in reality since a number of other dephasing mechanisms affect the spin dynamics and restrict the experimentally attained coherence time by a few seconds \cite{Zhan2018}. Below we clarify these mechanisms and show how their negative role could be partly overcome.

The main mathematical object of our consideration is the density operator $\hat{\rho}=\hat{\rho}(t)$ for a single atom confined with a particular trap, which implies quantum description of both the spin and spatial degrees of freedom of the atom.  It obeys the following master equation
\begin{equation}
\frac{d\hat{\rho}}{d t} = -\frac{i}{\hbar} \left[\hat{H}_{0},\,\hat{\rho}(t)\right] -\frac{i}{\hbar} \left[\hat{H}_{\mathrm{mw}}(t),\,\hat{\rho}(t)\right] +\left( \frac{\partial\hat{\rho}}{\partial t}\right)_{\mathrm{rel}},
\label{2.2}
\end{equation}
where the trap Hamiltonian $H_{0}$ is responsible for the natural dynamics of the atom inside the trap. The second term describes the interaction with the external microwave field, which is treated classically and assumed to be switched on for short durations at specific time moments. The pure hamiltonian dynamics is violated by the relaxation symbolically introduced by the last term in (\ref{2.2}) expressing the Lindblad-type transform of the density operator disturbed by the environment. Physically the interaction with the environment includes the stochastic disturbance of the trapping potential, induced by the fluctuations of the laser field confining the atom, and by the random process of incoherent scattering of this field by the atom.

\subsection{The trap Hamiltonian}

\noindent The alkali-metal atom, having high oscillator strength for a single valence electron, can be effectively coupled with a laser field even in a far off-resonant spectral domain. This coupling induces ac-Stark shifts of the atomic energy levels, which depend on the atom's location, with a minimum at the central point of the laser beam and creates a certain potential well restricting its spatial motion. If the initial kinetic energy of the atom is low enough it will experience an additional mechanical force originating in this barrier and can be trapped by such an optical tweezers system. Eventually the motion of the atom can be visualized as periodic oscillations inside the trap.

To describe such mechanical motion we have to take into consideration the fact that the atomic polarizabilities depend on the atomic spin variables because of hyperfine interactions. As a consequence the coupling of the atom with a far off-resonance trapping light can be described by means of adiabatic dressed-state-dependent forces, see  \cite{Dalibard1985}. In the subspace of the $|a\rangle$ and $|b\rangle$ spin states the undisturbed dynamics of the atom inside the trap is described by the following effective Hamiltonian
\begin{eqnarray}
H_{0}&=&-\frac{\hbar^2}{2m}\triangle +U_{a}(\hat{\mathbf{r}})|a\rangle\langle a|+ \left(\hbar\omega_{\mathrm{hpf}}+U_{b}(\hat{\mathbf{r}})\right)|b\rangle\langle b|%
\nonumber\\%
&&+U_{ab}(\hat{\mathbf{r}})|a\rangle\langle b|+ U_{ba}(\hat{\mathbf{r}})|b\rangle\langle a|%
\label{2.3},
\end{eqnarray}
where by putting a ``hat'' over the atom's position $\hat{\mathbf{r}}$ we emphasize its operator nature in the consistent quantum description. For the dipole trap prepared with the linearly polarized light the off-diagonal terms $U_{ab}(\hat{\mathbf{r}})$ and $U_{ba}(\hat{\mathbf{r}})$ are extremely small and can be safely neglected, which justifies the above pointed adiabatic approximation. So for the defined spin states the spatial motion of the atom is driven by two potentials
\begin{eqnarray}
U_{a}(\hat{\mathbf{r}})&\equiv&U_{0}(\hat{\mathbf{r}})+\hbar\Delta_{a}(\hat{\mathbf{r}})%
\nonumber\\%
U_{b}(\hat{\mathbf{r}})&\equiv&U_{0}(\hat{\mathbf{r}})+\hbar\Delta_{b}(\hat{\mathbf{r}})%
\label{2.4}
\end{eqnarray}
where we have specified the mean profile of the trap potential $U_{0}(\hat{\mathbf{r}})$ and linked the deviations with the additional light shifts $\Delta_{a}(\hat{\mathbf{r}})$ and $\Delta_{b}(\hat{\mathbf{r}})$ of the spin energy levels.

If the kinetic energy of the trapped atom is much smaller than the potential depth then near the bottom of the potential well the profiles can be approximated by a parabolic shape. In this case the Hamiltonian (\ref{2.3}) has the following sets of eigenfunctions
\begin{eqnarray}
|\mathit{v}_x\mathit{v}_y\mathit{v}_z;a\rangle&=&\psi_{\mathbf{v}}^{(a)}(\mathbf{r})|a\rangle\equiv |\psi_{\mathbf{v}}^{(a)};a\rangle\equiv |\mathbf{v};a\rangle%
\nonumber\\%
|\mathit{w}_x\mathit{w}_y\mathit{w}_z;b\rangle&=&\psi_{\mathbf{w}}^{(b)}(\mathbf{r})|b\rangle\equiv |\psi_{\mathbf{w}}^{(b)};b\rangle\equiv |\mathbf{w};b\rangle%
\label{2.5}
\end{eqnarray}
where for a sake of convenience we have denoted the sets of vibrational numbers in the potentials $U_{a}(\mathbf{r})$ and $U_{b}(\mathbf{r})$ by symbolic vectors $\mathbf{v}\equiv(\mathit{v}_x,\mathit{v}_y,\mathit{v}_z)$ and $\mathbf{w}\equiv(\mathit{w}_x,\mathit{w}_y,\mathit{w}_z)$, respectively. The vibrational functions $\psi_{\mathbf{v}}^{(a)}(\mathbf{r})$ and $\psi_{\mathbf{w}}^{(b)}(\mathbf{r})$, considered in the position representation, obey the Schr\"{o}dinger equations
\begin{eqnarray}
\left[-\frac{\hbar^2}{2m}\triangle + U_{a}(\mathbf{r})\right]\psi_{\mathbf{v}}^{(a)}(\mathbf{r})&=&\epsilon_{\mathbf{v}}\psi_{\mathbf{v}}^{(a)}(\mathbf{r})%
\nonumber\\%
\left[-\frac{\hbar^2}{2m}\triangle + U_{b}(\mathbf{r})\right]\psi_{\mathbf{w}}^{(b)}(\mathbf{r})&=&\varepsilon_{\mathbf{w}}\psi_{\mathbf{w}}^{(b)}(\mathbf{r})%
\nonumber\\%
\label{2.6}
\end{eqnarray}
where $\epsilon_{\mathbf{v}}\equiv\epsilon_{\mathit{v}_x\mathit{v}_y\mathit{v}_z}$ and $\varepsilon_{\mathbf{w}}\equiv\varepsilon_{\mathit{w}_x\mathit{w}_y\mathit{w}_z}$ are the vibrational energies in the potentials $U_{a}(\mathbf{r})$ and $U_{b}(\mathbf{r})$ respectively. Each of these equations generates a complete set of mutually orthogonal and normalized functions, but there is no orthogonality between the sets, such that $\langle \psi_{\mathbf{w}}^{(b)}|\psi_{\mathbf{v}}^{(a)}\rangle\neq 0$ for $\mathbf{w}\neq \mathbf{v}$.

Let us make the following remark important for our further derivation scheme. For eigenfunctions of harmonic oscillators and for the realistic parameters associated with our setup we have verified that $\langle \psi_{\mathbf{w}}^{(b)}|\psi_{\mathbf{v}}^{(a)}\rangle\approx 0$ if $\mathbf{w}\neq \mathbf{v}$ and $\langle \psi_{\mathbf{w}}^{(b)}|\psi_{\mathbf{v}}^{(a)}\rangle\approx 1$ if $\mathbf{w}= \mathbf{v}$ with very high accuracy. So for most of the further transformations we can ignore the non-orthogonality between the eigenfunctions of different sets. However the difference between the energies $\hbar\delta\omega_\mathbf{v} =  \left.\varepsilon_{\mathbf{w}}\right|_{\mathbf{w}=\mathbf{v}}-\epsilon_{\mathbf{v}}$ is not negligible and is a crucially important parameter for correct description of the decoherence process.

\subsection{The qubit control by a microwave field}

\noindent The on-resonant interaction of the atom with the microwave radiation is given by the following Hamiltonian
\begin{equation}
\hat{H}_{\mathrm{mw}}(t)=-\frac{\hbar}{2}\Omega(t)\mathrm{e}^{-i\omega t}|b\rangle\langle a| - \frac{\hbar}{2}\Omega^{\ast}(t)\mathrm{e}^{+i\omega t}|a\rangle\langle b|,%
\label{2.7}
\end{equation}
where $\omega$ is a carrier frequency and $\Omega=\Omega(t)$ is the time dependent Rabi frequency. Following the concept of the Hadamard gates, the qubit can be prepared by a $\pi/2$-pulse excitation. Then its coherent dynamics can be controlled via observation of the Ramsey resonance with a second $\pi/2$-pulse transferring the system to the state opposite to the initial one. Alternatively, the evidence of coherency can be justified by Rabi oscillations of the occupation probabilities of either $|a\rangle$ or $|b\rangle$ states once the stationary microwave field is turned on at an initial time, as shown in Fig.~\ref{fig1}(c).

For implication to the quantum data processing one has to prefer the former scenario with the pulse excitation and free qubit precession. 
In this particular case even for collection  of relatively warm atoms their mutual decoherence could be essentially suppressed with the use of the spin-echo protocol. We clarify such an option in Appendix \ref{Appendix_A} and further confirm it by our experimental data.

We can point out an important property of the excitation process adjusted by an infinitely short $\delta$-type mw-pulse. Suppose that before excitation the atom occupies any spin state and exists in an arbitrary vibrational state $|\psi_{\mathrm{vib}}\rangle$. Then, as shown in Appendix \ref{Appendix_A} this vibrational state cannot be changed by chirping the system by the spectrally broad mw-pulse having short duration, such that
\begin{eqnarray}
\dot{c}_{b}(t)|\psi_{\mathrm{vib}}\rangle&=&\frac{i}{2}\,\Omega(t)\,c_{a}(t)\,|\psi_{\mathrm{vib}}\rangle%
\nonumber\\%
\dot{c}_{a}(t)|\psi_{\mathrm{vib}}\rangle&=&\frac{i}{2}\,\Omega^{\ast}(t)\,c_{b}(t)\,|\psi_{\mathrm{vib}}\rangle%
\label{2.8}%
\end{eqnarray}
where $c_{a}(t)$ and $c_{b}(t)$ are the probability amplitudes for the spin wavefunction superposed between $|a\rangle$ and $|b\rangle$. So the spin and vibrational subsystems stay disentangled after such a short pulse excitation. It makes difference if the atom initially exists in an entangled composition of the spin and vibrational degrees of freedom. Then, as shown in Appendix \ref{Appendix_A}, the short $\pi$-pulse initiates a flip of the spin and vibrational states of the atom. This option gives us an example of the spin-echo protocol and can be used to reverse the system dynamics towards revival of its initial state.

\subsection{Interaction with environment}\label{Subsection_II.D}

\noindent The evolution of an individual atom in a far off-resonant optical dipole trap (optical tweezers) is described by a Schr\"{o}dinger equation with the following extra term added to the atomic Hamiltonian
\begin{equation}
\hat{H}_{\mathrm{eff}}=\sum_{n,m',m}\frac{(\hat{\mathbf{d}}\cdot\hat{\mathbf{E}}^{(-)}(\mathbf{r}))_{m'n}(\hat{\mathbf{d}}\cdot\hat{\mathbf{E}}^{(+)}(\mathbf{r}))_{nm}}{\hbar\left(\omega_L-\omega_{nm}\right)}\,%
|m'\rangle\langle m|.%
\label{2.9}
\end{equation}
This correction occurs due to the dipole interaction contributing in the second order of the perturbation theory, see \cite{Happer1972,Mishina2005}. Here $\hat{\mathbf{d}}$ is the atomic dipole moment operator and $\hat{\mathbf{E}}^{(+)}(\mathbf{r})$ and $\hat{\mathbf{E}}^{(-)}(\mathbf{r})$ are the operators of the positive and negative frequency components of the field taken at the position of the atom. Under steady state conditions the Hamiltonian can be approximated by its mean value such that the field operators can be substituted by the complex amplitudes of the laser field $\mathbf{E}_0(\mathbf{r})$ and $\mathbf{E}_0^{\ast}(\mathbf{r})$, which in a paraxial limit have a Gaussian mode profile. That simplifies the problem to coherent coupling of the atom with the laser mode and we get
\begin{eqnarray}
\lefteqn{\hspace{-2cm}\sum_{n,m',m}\frac{(\hat{\mathbf{d}}\cdot\mathbf{E}_0^{\ast}(\mathbf{r}))_{m'n}(\hat{\mathbf{d}}\cdot\mathbf{E}_0(\mathbf{r}))_{nm}}{\hbar\left(\omega_L-\omega_{nm}\right)}\,|m'\rangle\langle m|}
\nonumber\\%
&\equiv&\sum_{n,m',m}U_{m'm}(\mathbf{r})\,|m'\rangle\langle m|.%
\label{2.10}
\end{eqnarray}
For the far off-resonant laser frequency $\omega_L$ the denominator in (\ref{2.10}) becomes insensitive to the upper state energy structure and eventually transforms to
\begin{equation}
U_{m'm}(\mathbf{r})\approx U_{m}(\mathbf{r})\delta_{m'm},%
\label{2.11}
\end{equation}
where $U_{m}(\mathbf{r})$ with $m=a,b$ define the trap potentials $U_{a}(\hat{\mathbf{r}})$ and $U_{b}(\hat{\mathbf{r}})$ with the position variable treated as an operator in Eqs.~(\ref{2.3}), (\ref{2.4}).

In reality the above arguments are valid only approximately since the electromagnetic field has an intrinsically stochastic nature. There are two basic physical mechanisms of interaction with the environment which have to be taken into consideration. First, the laser radiation is not stable. The profile potential will fluctuate randomly due to the intensity fluctuation in the trapping light beam, \cite{Savard1997}. Even if it is stable from the classical point of view it will possess the quantum fluctuations in the photon flux at the Poissonian level of uncertainty. In our numerical simulations we consider classical fluctuations, which are described as a classical Wiener-type stochastic process. We used the experimentally measured values of relative intensity noise, which in our case is much larger than the shot noise, justifying its classical treatment, see Appendix \ref{Appendix_B} for details. The potentially interesting quantum effects associated with the sub-Poisson statistics or light squeezing are beyond our consideration here. Second, the laser light can be incoherently scattered from the atom via Raman scattering channels \cite{Heinzen1994}.\footnote{As we comment in Appendix \ref{Appendix_B} the Rayleigh scattering preserves the spin coherence.} These two mechanisms disturb the mechanical confinement of the atom with the profile potentials (\ref{2.10}) and have to be added to the Hamiltonian for treating the atom as an open quantum system.

The deviation (\ref{2.9}) from (\ref{2.10}) incorporates both processes
\begin{eqnarray}
\hat{W}&=&\hat{H}_{\mathrm{eff}}-\sum_{n,m',m}U_{m'm}(\mathbf{r})\,|m'\rangle\langle m|%
\nonumber\\%
&\approx&\sum_{n,m}\left[\frac{(\hat{\mathbf{d}}\cdot\hat{\mathbf{E}}^{(-)}(\mathbf{r}))_{mn}(\hat{\mathbf{d}}\cdot\hat{\mathbf{E}}^{(+)}(\mathbf{r}))_{nm}}{\hbar\left(\omega_L-\omega_{0}\right)}%
-U_0({\mathbf{r}})\right]%
\nonumber\\%
\nonumber\\%
&&\phantom{\sum_{n,m',m}U_{m'm}(\mathbf{r})\,|m'\rangle\langle m|}\times|m\rangle\langle m|,%
\label{2.12}%
\end{eqnarray}
where in the second equality we have assumed that for this perturbation term and in Eq.~(\ref{2.11}) any relativistic effects for the atomic valence electron can be ignored. Such a simplification is justified by the considered far off-resonant conditions when an alkali-metal atom can be approximated by a two-level atom with the mean transition frequency $\omega_0=(2\omega_2+\omega_1)/3$, where $\omega_1$ and $\omega_2$ are the respective frequencies of $D_1$ and $D_2$ lines of the atom. That, in turn, is justified by the inequality $|\omega_L-\omega_0|\gg\omega_2-\omega_1$ which is satisfied under our experimental conditions. The relativistic effects are very important for the correct description of the evolution of the trapped atom driven by the microwave radiation, but they can be safely neglected in the estimate (\ref{2.12}). Thus (i) the interaction Hamiltonian (\ref{2.12}) is diagonal and independent on the spin variables in the basis (\ref{2.5}), and (ii) for any of its matrix elements there is no difference between vibrational wavefunctions (\ref{2.6}) belonging to different spin states, such that the trap potential can be approximated by its mean profile $U_0(\mathbf{r})$.

\section{Experiment}\label{Section_III}

In the experiment, the microscopic dipole trap is formed by a beam of an 852~nm diode laser, tightly focused with an 0.77 NA aspheric lens installed inside a UHV chamber with a base pressure below $10^{-10}$~mbar. A detailed description of the setup may be found in \cite{Samoylenko_LPL2020}. Focusing of the trapping laser to a $1/{\rm e}^2$ intensity waist around 1.4~$\mu$m allows us to trap atoms ensuring single atom occupation by collisional blockade in the presence of MOT light \cite{Grangier_Nature2001}. The waist size was estimated from the analysis of vibrational frequencies of the trap. The values of the longitudinal and transverse vibrational trap frequencies were found to be 9.6~kHz and 72~kHz, respectively. The shape and position of the traps are controlled by a spatial light modulator, which is also used to control the beam power and therefore the trap depth. The effective temperature of the trapped atoms is determined by probing the energy distribution of the trapped atom with an adiabatic trap depth lowering method, described in \cite{Grangier_PRA2008}. In this method the trap depth is slowly reduced to some value $U_{\mathrm{min}}$ then brought back to its initial value, and the probability for the atom to remain in the trap is measured. Fitting the dependence of this probability on $U_{\mathrm{min}}$ with a simple model assuming initially thermal motional state of the atom and adiabaticity results in an estimate of temperature. In our case the temperature is measured to be 40$\pm$7~$\mu$K under typical experimental conditions. This temperature level corresponds to the typical values which can be attained via the polarization gradient cooling mechanism in MOT. Further cooling is a next step, and in our future work we are planning to slow down the atomic vibrational motion with RSC protocol.

The experimental sequence starts by optically pumping the atom to the $|a\rangle = |F_-=1, M_0=0\rangle$ with a $\pi$-polarized laser at the $|F_-=1\rangle \rightarrow |F'=1\rangle$ of the D1 line acting together with a cooling laser in a presence of a $1.8$~Gauss bias magnetic field collinear with the dipole trap axis. We estimate the efficiency of pumping to a particular hyperfine manifold of 95\% with push-out measurements. The push-out technique allows to implement hyperfine state selective measurements and is performed by applying a strongly focused circularly polarized push-out beam at the $|F_+=2\rangle \rightarrow |F'=3\rangle$ closed cycling transition, removing the atom from the trap if it is in the $F_+=2$ manifold. The action of the push-out beam is followed by exciting fluorescence with the MOT beams which is detected with an sCMOS camera (Tucsen Dhyana 400 bsi v2). Successful fluorescence detection indicates the atomic state $F_-=1$ prior to push-out, thus the whole sequence corresponds to a projective measurement of the hyperfine state.

After the initial state $|a\rangle = |F_{-}=1, M_{0}=0\rangle$ is prepared by optical pumping, we apply a linearly polarized radiofrequency (RF) field with an antenna installed outside the vacuum chamber. Let us note, that due to the presence of metallic parts of the lens holder inside the vacuum chamber, which form a subwavelength aperture for the RF radiation, the RF field polarization at the position of the atom is unknown. We drive the clock transition with the Rabi frequency of 1.5~kHz and vary the pulse sequences as required for Ramsey or spin-echo interferometry. The corresponding RF pulse sequences are shown in Fig.~\ref{fig2} (a) and Fig.~\ref{fig3} (a), respectively.  Finally, the projective measurement in the logical basis is performed using the push-out technique.  


\section{Results}\label{Section_IV}

\begin{figure}[t]
\scalebox{0.5}{\includegraphics*{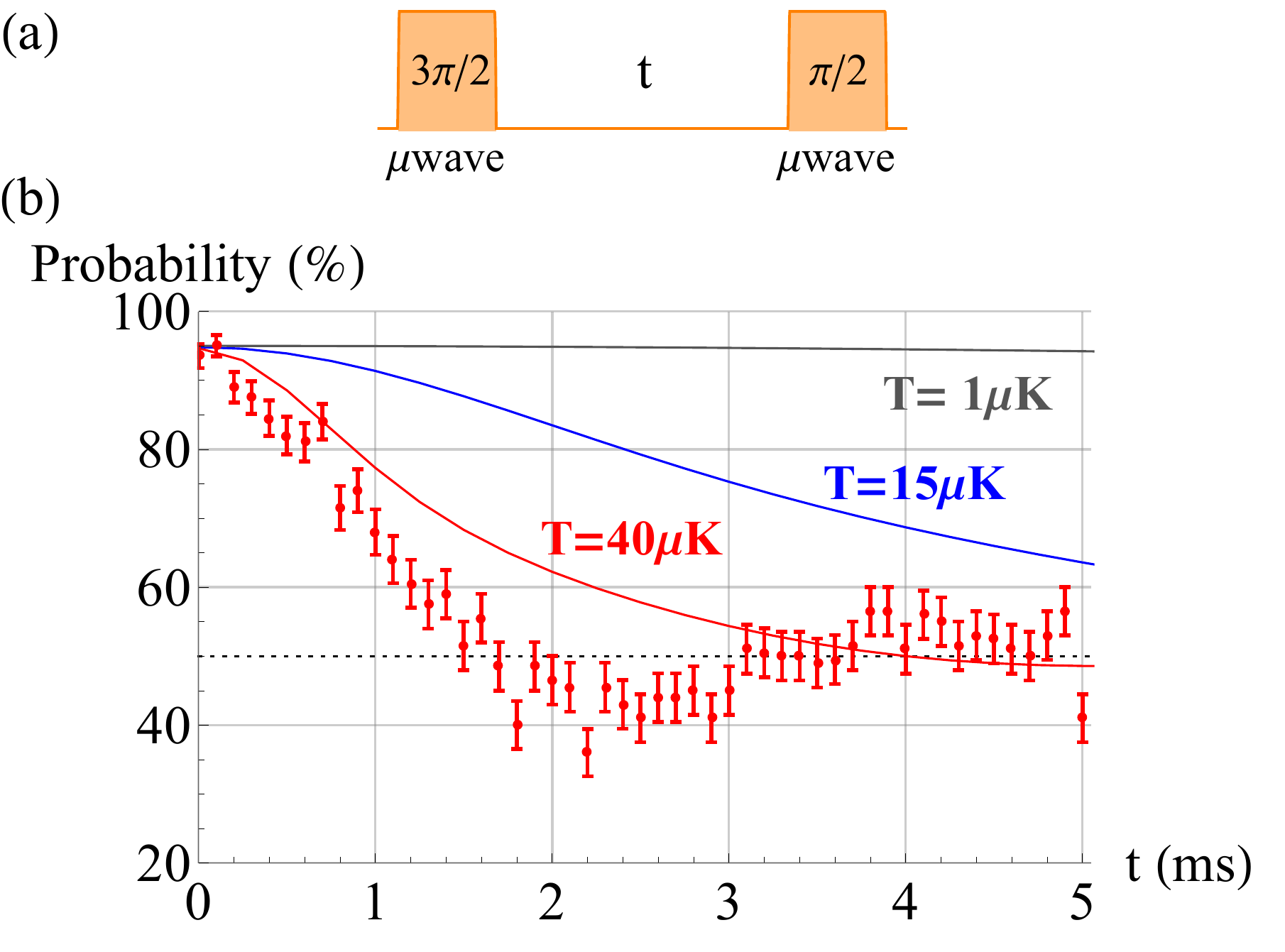}}
\caption{(Color online) (a) Pulse sequence, (b) Population of the $|a\rangle$-state recorded by the Ramsey resonance (points with error bars) vs. its theoretical estimate (solid curves). The slight oscillations on the experimental dependence indicate the residual imperfection of  the microwave pulses, see text. The curve colour is associated with the temperature of a trapped atom $T = 1,\, 15,\, 40\,\mu$K.}
\label{fig2}%
\end{figure}

\noindent In this section we present the results of our numerical simulations and compare them with the experimental data. The measurement protocol is based on observation of the Ramsey resonance, when the qubit dynamics is initiated and controlled by a sequence of two short microwave pulses of $3\pi/2$- and $\pi/2$-types delayed in time, such that in the ideal scenario the microwave radiation induces the population transfer to the initial state $|a\rangle$. In Fig.~\ref{fig2} we show the response signal, up to a given time $t$, in the population of the initial atomic state, which was experimentally measured and independently calculated numerically. The fidelity of the stored qubit, expressed by efficiency of recovering the original atomic state, degrades from its preparation level within a few  milliseconds. This kind of coherence damping is associated with the inhomogeneous dephasing under the internal system dynamics without environment losses. The observed dynamics is primarily driven by the Hamiltonian ${\hat H}_0 + {\hat H}_{\rm mw}(t)$, and the dephasing time in Fig.~\ref{fig2} can be scaled by the inverse value of the mean difference between vibrational frequencies belonging to the different basis spin states $\langle\delta\omega_{\mathbf{v}}\rangle\propto T$ (see (\ref{a.7})).

For unambiguous comparison with the measured data (points with error bars) in Fig.~\ref{fig2} we present our calculations (solid lines) reproducing the exact measurement conditions with varying the temperature of the thermal distribution. The temperature of a loaded atom was independently measured in the experiment and the presented data gives an unbiased theoretical estimate of the qubit characteristics with no fitting parameters. Let us note that a large number of resonant transitions arise for a relatively warm atom driven by a spectrally broadband pulse. The process of population transfer was optimized by microwave carrier frequency $\omega$ scanning to maximize the observed Rabi oscillations contrast (see Fig.~\ref{fig1}(c)). A similar procedure was followed to find the optimal carrier frequency $\omega$ in our numerical simulations. The imperfections of this method results in a slight signature of time beats in the experimental data shown in Fig.~\ref{fig2}.

The strong influence of the atomic motion on the spin dynamics makes the detection scheme sensitive to the duration parameters of the microwave pulses. In an ideal scenario they have to be infinitely short such that their spectral band would significantly overlap with the mean difference between the energies of the excited vibrational modes for a given temperature. For a relatively warm atom this condition cannot be perfectly fulfilled, since the respective frequency mismatch $\langle\delta\omega_{\mathbf{v}}\rangle\propto T$ is increasing with the temperature $T$. However, we observe a negligible imperfection of the Ramsey protocol with resonant microwave pulses with duration of $\tau_{\rm mv} \sim 100\ \mu{\rm s}$ and for the atom prepared at temperatures of tens $\mu$K, such that $\tau_{\rm mv}^{-1} \gg \langle\delta\omega_{\mathbf{v}}\rangle$. Our calculations of the Ramsey signal were done for the $\pi/2$-pulses with finite duration as well as for their infinitely short $\delta$-type approximations. There is negligible difference between these two theoretical estimates within the graph resolution.


The decoherence process, associated with the thermal motion of the atom, has a clear signature of non-exponential behavior. That is confirmed by our experimental data as well as by our theoretical estimates. Let us note, that the conventional dephasing time parameter ``$T_2^{\ast}$'' can only be introduced in the case of exponential decay of the Ramsey fringes. So it is of limited value here and cannot realistically capture some features of the observed behaviour. The theory predicts even more --- for longer time and in the absence of other relaxation processes, besides the inhomogeneous dephasing due to the thermal motion, the qubit has an ability for revival of its initial state. Unfortunately for the attained temperatures this would happen at the waiting time much longer than the unavoidable homogeneous decoherence caused by environment losses. But if the trapped atom was slowed down to its ground vibrational mode the process of inhomogeneous relaxation could be completely eliminated.

\begin{figure}[t]
\scalebox{0.5}{\includegraphics*{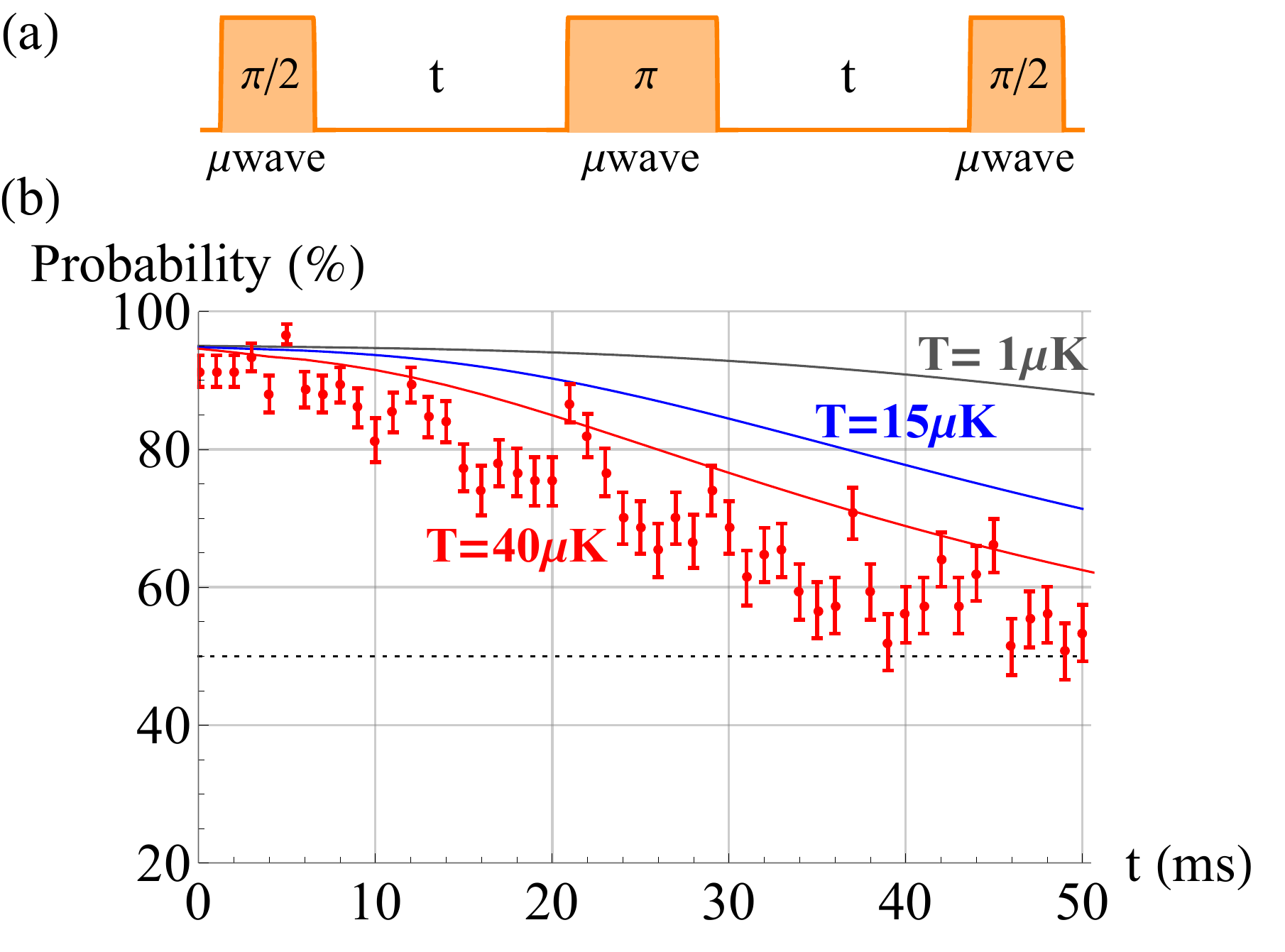}}
\caption{(Color online) (a) Pulse sequence, (b) Same as in Fig.~\ref{fig2} but measured with the spin-echo detection protocol. We insert a $\pi$-pulse in the middle of the gap between the $\pi/2$-control pulses, such that time $t = t_{\pi} - t_{\pi/2} = t'_{\pi/2} - t_{\pi}$, where $t_{\pi/2}$, $t_\pi$, $t'_{\pi/2}$ are the arrival times of the pulses. 
}
\label{fig3}%
\end{figure}

Intuitively it might seem that the problem with the atomic motion could be resolved by applying a spin-echo detection protocol. However the potential of this option has to be critically analyzed in context of qubit preparation as an element of a quantum register. The spin echo concept implies an idea of time reversal symmetry and suggests an additional excitation of the system by a short $\pi$-type pulse just in the middle point of time when the Ramsey resonance would be detected. Then up to this detection time the qubit accumulates an equal phase independently on its vibrational mode. In Fig.~\ref{fig3} we show the signal of the Ramsey resonance detected with the spin-echo protocol which indicates the significantly improved coherency in the qubit dynamics.

As in previous case,  in experiment, the control pulses excite the atom resonantly. 
Both the experimental data and theoretical simulations demonstrate non-exponential decay. In the supplementary materials, see Appendix \ref{Appendix_B}, we explain it and describe two physical mechanisms of the homogeneous decoherence associated with (i) the intensity fluctuations and (ii) incoherent scattering of the trapping light.

\begin{figure}[t]
\scalebox{0.5}{\includegraphics*{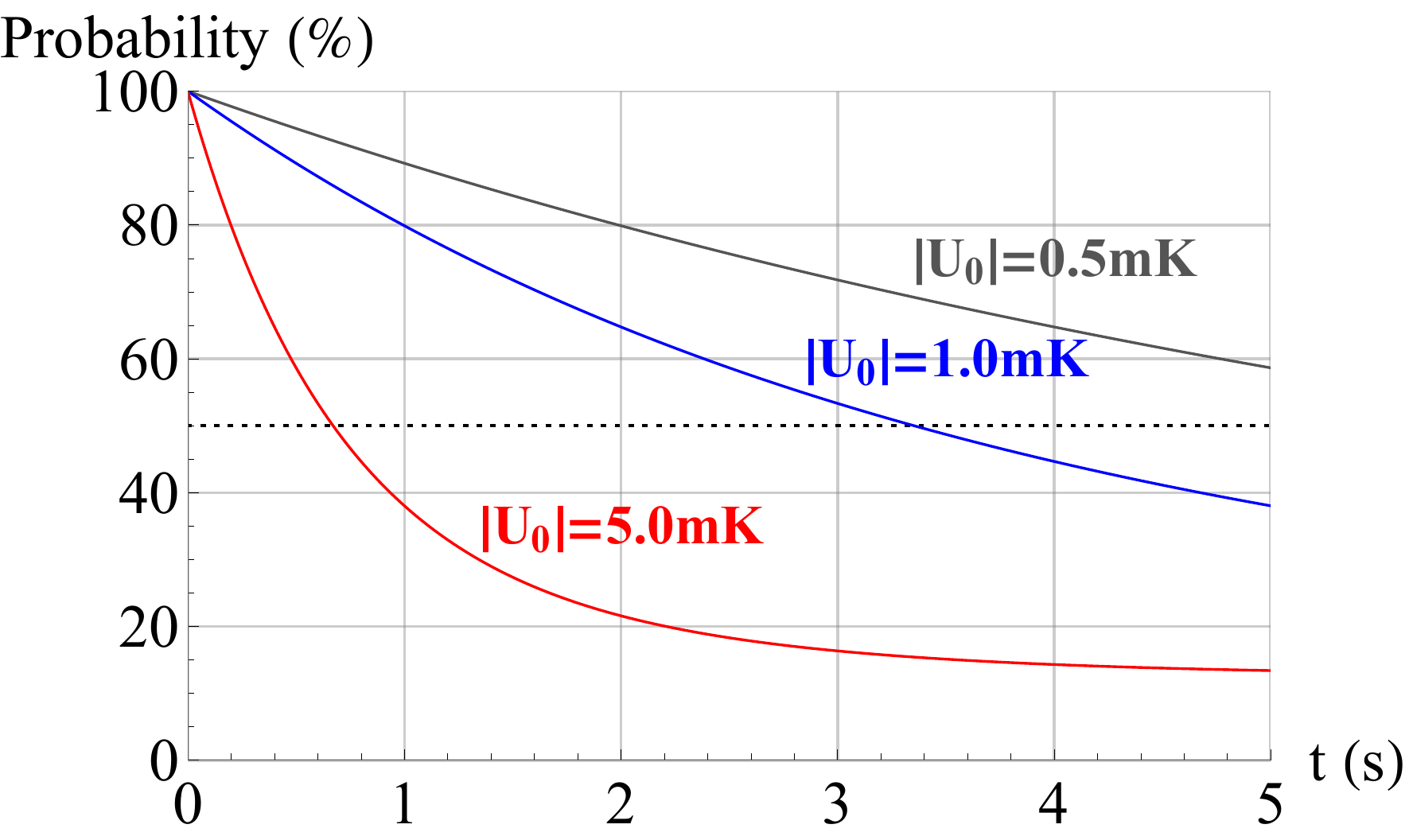}}
\caption{(Color online) Ramsey signal calculated for an atom at zero temperature in a Lamb-Dicke regime, in which its motional state does not change in the process of trap photons scattering. We set the carrier frequency in resonance with shifted hyperfine transition: $\omega = \omega_{\rm hpf} + \delta\omega_\mathbf{0}$, see (\ref{a.7}). The signals are plotted for trap depths of $|U_0| \simeq 0.5\,$mK (gray),  $|U_0| \simeq 1\,$mK (blue) and $|U_0| \simeq 5\,$mK (red). In lower trap depth the photon scattering rate decreases and spin coherence is shown to degrade slower.}
\label{fig4}%
\end{figure}

The specific dependencies in Fig.~\ref{fig4}, calculated for zero temperature, predict that with slowing the atom down to its ground vibrational state and for a tight dipole trap, having a small Lamb-Dicke parameter, both these mechanisms could be strongly suppressed. In this case the coherence demolishing would be limited by relativistic corrections in the scattering process of the trapping light and the prepared qubit could remain coherent for a few seconds \cite{Heinzen1994,Meshede2000}.

Note that in Fig.~\ref{fig4} longer decay times were obtained in lower trap depths $|U_0|$ ($U_0 = U_{a}({\mathbf{r}}=0)\approx U_{a}({\mathbf{r}}=0)$), since the atom in a red-detuned dipole trap is localized in an intensity maximum and thus higher scattering rate of trap photons occurs for higher $|U_0|$. Reaching the strong Lamb-Dicke regime in a shallow trap with small $|U_0|$ can be problematic. However one may hope that both requirements of small Lamb-Dicke parameter and low scattering rate can be met in a blue-detuned trap, where the potential depth $U_0 > 0$ is determined by the height of the repulsive walls surrounding the center of the trap \cite{Grimm2000}. In this case a potential minimum corresponds to an intensity minimum, which in an ideal case means zero intensity, hence, negligible homogeneous dephasing due to trap photons scattering.

An important consequence of such optimal conditions is that there would be no need in making use of the spin-echo protocol since the inhomogeneous dephasing would vanish in this case. We would also point out that the spin-echo protocol seems to be not so convenient for manipulation with a multi-qubit system, which implies the controllable entanglement of particular qubits in the quantum register at intermediate stages of the entire cycle of quantum simulations.

The schemes of Raman sideband cooling (RSC) adjusted for the optical tweezers systems give us a tool for further optimization, see \cite{RSC2019,Andersen2017}. To make the most effective use of RSC protocol and reach an ideal three-dimensional ground state cooling one needs to meet the requirements of strong Lamb-Dicke regime. For typical parameters of a dipole trap with axial oscillator frequency $\omega_\parallel \ll T$ the mean value of vibrational number can be estimated as $\bar{v}_\parallel \sim 100$. This can be an important constraint for the RSC protocol that requires additional optimization of its parameters during the cooling procedure. However, there are certain difficulties for experimental preparation of a dipole trap, which would be sufficiently tight in axial direction. Hence a two-dimensional scheme of RSC with the cooling of transverse vibrational atomic motion only is more convenient to implement in a real experimental setup. Since the temperature is a critical parameter of the system, even suppressing transverse vibrational motion only may lead to a notable improvement of coherence time and may allow us to resolve the motional decoherence problem in a multiatomic system.

\section{Conclusion}

In the paper we have presented a theoretical analysis and numerical simulation which treat the various decay mechanisms of the hyperfine coherence of an optically trapped neutral atom. The realistic modeling of dephasing processes due to the atomic residual motion, trap light intensity fluctuations and incoherent scattering has been performed. The numerical results reliably reproduce the experimentally observed shapes of Ramsey and spin-echo signals with no fitting parameters.

Our theoretical and experimental observations have confirmed the notable improvement of coherency in the qubit dynamics detected using the spin-echo technique. However, certain homogeneous dephasing mechanisms still limit the coherence time measured with the spin-echo detection protocol. As we have shown and highlighted throughout our discussion, to achieve significantly longer coherence times one has to quench the motion of an atom down to its vibrational ground state in the potential well and further optimize the parameters of the optical dipole trap. This would imply sufficiently tight confinement and trapping of atoms at the minimum of electric fields.

In our numerical simulations we treat the atomic motional degrees of freedom quantum mechanically, which allows us to correctly describe a weakly excited atomic oscillator.
It was shown that the coherence time of atomic qubit having a zero temperature in a tight dipole trap under conditions of strong Lamb-Dicke regime is limited by the scattering process accompanied by the Raman transitions, changing the atomic spin states, which occur only through the relativistic fine-structure interaction. This process is strongly reduced in the case of a far detuned optical dipole trap and thus the coherence decay time exceeding several seconds could be achieved. Experimental exploration of this regime is an important topic for future work.

\acknowledgments
\noindent This work was supported by the Russian Foundation for Basic Research under Grants \# 18-02-00265-A and \#19-52-15001-CNRS-a, by the Russian Science Foundation under Grant \# 18-72-10039, and by the Foundation for the Advancement of Theoretical Physics and Mathematics ``BASIS'' under Grant \# 18-1-1-48-1. R.R.~Yu. and D.~Sh. acknowledge support from Foundation for Assistance to Small Innovative Enterprises under Grant UMNIK.

\appendix
\section{Coherent dynamics of the trapped atoms driven by the microwave radiation}\label{Appendix_A}
\noindent In order to find the solution of the dynamical part of Eq.~(\ref{2.2}) only, without loss of generality, it is sufficient to construct it for the separable density operator factorized to a projector onto a specific pure state described by a time-dependent wavefunction. Its dynamical evolution involves any pairs of the vibrational states associated with $|a\rangle$ and $|b\rangle$ and can be expressed by the sets of probability amplitudes
\begin{eqnarray}
\hat\rho(t)&=&|\Psi(t)\rangle\langle\Psi(t)|%
\nonumber\\%
\nonumber\\%
|\Psi(t)\rangle&=&\sum_{\mathbf{v,w}}\left[c_{a,\mathbf{v}}(t)\,\mathrm{e}^{-\frac{i}{\hbar}\epsilon_{\mathbf{v}}t}\,|\psi_{\mathbf{v}}^{(a)};a\rangle\right.%
\nonumber\\%
&&\left.+\,c_{b,\mathbf{w}}(t)\,\mathrm{e}^{-i\omega_{\mathrm{hpf}}t-\frac{i}{\hbar}\varepsilon_{\mathbf{w}}t}\,|\psi_{\mathbf{w}}^{(b)};b\rangle\right]\, ,%
\label{a.1}%
\end{eqnarray}
which obey the Schr\"{o}dinger equation written in the energy representation
\begin{eqnarray}
\dot{c}_{b,\mathbf{w}}(t)&=&\frac{i}{2}\,\Omega(t)\sum_{\mathbf{v}}\langle\psi_{\mathbf{w}}^{(b)}|\psi_{\mathbf{v}}^{(a)}\rangle\,%
\mathrm{e}^{-i\left(\omega -\omega_{\mathrm{hpf}}-\omega_{\mathbf{w}\mathbf{v}}\right)t}\,c_{a,\mathbf{v}}(t)%
\nonumber\\%
\dot{c}_{a,\mathbf{v}}(t)&=&\frac{i}{2}\,\Omega^{\ast}(t)\sum_{\mathbf{w}}\langle\psi_{\mathbf{v}}^{(a)}|\psi_{\mathbf{w}}^{(b)}\rangle\,%
\mathrm{e}^{i\left(\omega -\omega_{\mathrm{hpf}}-\omega_{\mathbf{w}\mathbf{v}}\right)t}\,c_{b,\mathbf{w}}(t).%
\nonumber\\%
\label{a.2}%
\end{eqnarray}
It is an intriguing point that the expansion (\ref{a.1}), considered at an arbitrary moment of time, is an entangled state even if the dynamics is free and the spin state was initially disentangled with the vibrational motion. That is just the consequence of difference between the trap potentials $U_{a}(\hat{\mathbf{r}})$ and $U_{b}(\hat{\mathbf{r}})$. Nevertheless the atom could be prepared in one of the spin states and therefore would have a separate vibrational wavefunction $|\psi_{\mathrm{vib}}\rangle$.\footnote{One can imagine the atom to be prepared in a disentangled state between the spin and vibrational degrees of freedom $|s\rangle\otimes|\psi_{\mathrm{vib}}\rangle$ at zero time. But it is not an eigenstate of the Hamiltonian, so it would transform to an entangled state at arbitrary time.}

Let the atom be prepared in a spin state $|a\rangle$. If $\Omega(t)$ contributes as a short $\delta$-type pulse, then all the vibrational modes in the upper states can be equally excited. Due to completeness of the set of oscillator functions
\begin{equation}
\sum_{\mathbf{w}}\langle\psi_{\mathbf{w}}^{(b)}|\psi_{\mathbf{v}}^{(a)}\rangle\,|\psi_{\mathbf{w}}^{(b)}\rangle=|\psi_{\mathbf{v}}^{(a)}\rangle%
\label{a.3}%
\end{equation}
and to a similar relation with replacement $a\rightleftharpoons b$, we obtain that the excitation of the spin coherency does not influence the vibrational motion of the atom. Indeed, to verify this, one need to multiply the first equation of (\ref{a.2}) on $|\psi_{\mathbf{w}}^{(b)}\rangle$ and the second line on $|\psi_{\mathbf{v}}^{(b)}\rangle$ and add the sums over $\mathbf{w}$ and $\mathbf{v}$ respectively. So for an infinitely short pulse the equations can be rewritten as
\begin{eqnarray}
\dot{c}_{b}(t)|\psi_{\mathrm{vib}}\rangle&=&\frac{i}{2}\,\Omega(t)\,c_{a}(t)\,|\psi_{\mathrm{vib}}\rangle%
\nonumber\\%
\dot{c}_{a}(t)\,|\psi_{\mathrm{vib}}\rangle&=&\frac{i}{2}\,\Omega^{\ast}(t)\,c_{b}(t)\,|\psi_{\mathrm{vib}}\rangle%
\label{a.4}%
\end{eqnarray}
and coefficients $c_{a,\mathbf{v}}(t)$ and $c_{b,\mathbf{w}}(t)$ can be searched in the form $c_{a,\mathbf{v}}(t)=c_{a}(t)\,\alpha_{\mathbf{v}}$ and $\sum_{\mathbf{w}}c_{b,\mathbf{w}}(t)|\psi_{\mathbf{w}}^{(b)}\rangle=c_{b}(t)|\psi_{\mathrm{vib}}\rangle$, where
\begin{equation}
|\psi_{\mathrm{vib}}\rangle\equiv\sum_{\mathbf{v}}\alpha_{\mathbf{v}}\,|\psi_{\mathbf{v}}^{(a)}\rangle.%
\label{a.5}
\end{equation}
It makes evident that the system transforms to the conventional two-level problem  for any [conserved] vibrational state $|\psi_{\mathrm{vib}}\rangle$. As a consequence, an arbitrary collection of trapped atoms, even if they are mutually mobile, can be converted to the coherent register state given by the product of qubits via the Hadamard transform initiated by a short $\pi/2$ microwave pulse and this can be done with ideal fidelity.

In general case the system (\ref{a.2}) cannot be analytically solved. But it can be simplified and reliably approximated by a two-level system if we keep only the pairs of states with $\mathbf{w}=\mathbf{v}$ so that the evolution operator takes a form of a block matrix. For the each elementary block we have
\begin{eqnarray}
\dot{c}_{b,\mathbf{v}}(t)&\cong&\frac{i}{2}\,\Omega(t)\,%
\mathrm{e}^{-i\left(\omega -\omega_{\mathrm{hpf}}-\delta\omega_{\mathbf{v}}\right)t}\,c_{a,\mathbf{v}}(t)%
\nonumber\\%
\dot{c}_{a,\mathbf{v}}(t)&\cong&\frac{i}{2}\,\Omega^{\ast}(t)\,%
\mathrm{e}^{i\left(\omega -\omega_{\mathrm{hpf}}-\delta\omega_{\mathbf{v}}\right)t}\,c_{b,\mathbf{v}}(t),%
\nonumber\\%
\label{a.6}%
\end{eqnarray}
where we have assumed that $\langle\psi_{\mathbf{v}}^{(b)}|\psi_{\mathbf{v}}^{(a)}\rangle\simeq 1$ and denoted
\begin{eqnarray}
\lefteqn{\delta\omega_{\mathbf{v}}=\left(\left.\varepsilon_{\mathbf{w}}\right|_{\mathbf{w}=\mathbf{v}}-\epsilon_{\mathbf{v}}\right)/\hbar}%
\nonumber\\%
\nonumber\\%
&&=\left(\mathit{v}_x+\mathit{v}_y +1\right)\left[\omega_{\perp}^{(b)}-\omega_{\perp}^{(a)}\right]+(\mathit{v}_z+1/2)\left[\omega_{\parallel}^{(b)}-\omega_{\parallel}^{(a)}\right],%
\nonumber\\%
\label{a.7}
\end{eqnarray}
which clarifies the difference between the energies of the same vibrational excitations but in the different potentials. Here $\omega_{\perp}^{(a)}$, $\omega_{\perp}^{(b)}$ and $\omega_{\parallel}^{(a)}$, $\omega_{\parallel}^{(b)}$ are respectively the transverse and axial oscillators' frequencies in these potentials. The evolution operator (a fundamental solution of (\ref{a.6})) can be found in a closed form for the case of a rectangular pulse with $\Omega=\mathrm{const}$ within its duration and in that case is expressed by the following matrix:
\begin{widetext}
\begin{equation}
\hat{{\cal U}}_{\mathbf{v}}(t)=\left[\begin{array}{cc}\left[\cos\left(\displaystyle\frac{\Omega_{\mathbf{v}} t}{2}\right)%
+i\displaystyle\frac{\Delta_{\mathbf{v}}}{\Omega_{\mathbf{v}}}\sin\left(\displaystyle\frac{\Omega_{\mathbf{v}} t}{2}\right)\right]\,%
\mathrm{e}^{-i\Delta_{\mathbf{v}}t/2}%
&i\displaystyle\frac{|\Omega|}{\Omega_{\mathbf{v}}}\sin\left(\displaystyle\frac{\Omega_{\mathbf{v}} t}{2}\right)\,\mathrm{e}^{i\phi-i\Delta_{\mathbf{v}}t/2}\\ \\%
i\displaystyle\frac{|\Omega|}{\Omega_{\mathbf{v}}}\sin\left(\displaystyle\frac{\Omega_{\mathbf{v}} t}{2}\right)\,\mathrm{e}^{-i\phi+i\Delta_{\mathbf{v}}t/2}%
&\left[\cos\left(\displaystyle\frac{\Omega_{\mathbf{v}} t}{2}\right)-i\displaystyle\frac{\Delta_{\mathbf{v}}}{\Omega_{\mathbf{v}}}\sin\left(\displaystyle\frac{\Omega_{\mathbf{v}} t}{2}\right)\right]\,%
\mathrm{e}^{i\Delta_{\mathbf{v}}t/2}\end{array}\right],%
\label{a.8}
\end{equation}
\end{widetext}
where $\Delta_{\mathbf{v}}=\omega -\omega_{\mathrm{hpf}}-\delta\omega_{\mathbf{v}}$, $\Omega=|\Omega|\mathrm{e}^{i\phi}$ and we have defined the generalized Rabi frequency as
\begin{equation}
\Omega_{\mathbf{v}}=\sqrt{|\Omega|^2+\Delta_{\mathbf{v}}^2}.%
\label{a.9}%
\end{equation}
Then the solution of (\ref{a.6}) is given by
\begin{equation}
\left[\begin{array}{c}c_{b,\mathbf{v}}(t)\\c_{a,\mathbf{v}}(t)\end{array}\right]=\hat{{\cal U}}_{\mathbf{v}}(t)%
\left[\begin{array}{c}c_{b,\mathbf{v}}(0)\\c_{a,\mathbf{v}}(0)\end{array}\right].%
\label{a.10}%
\end{equation}
The complete evolutionary operator can be compiled as a block structured matrix
\begin{equation}
\hat{\boldsymbol{\cal U}}(t)=\left[\begin{array}{ccccc}\hat{{\cal U}}_{0,0,0}(t)&0&0&0&\ldots\\0&\hat{{\cal U}}_{1,0,0}(t)&0&0&\ldots\\0&0&\hat{{\cal U}}_{0,1,0}(t)&0&\ldots\\ 0&0&0&\hat{{\cal U}}_{0,0,1}(t)&\ldots\\%
\ldots&\ldots&\ldots&\ldots&\ldots\end{array}\right]%
\label{a.11}
\end{equation}
The tricky point is that each block in the spin subspace is associated with the dyadic operator in the vibrational subspace of the two potentials $U_{a}(\hat{\mathbf{r}})$ and $U_{b}(\hat{\mathbf{r}})$, parameterized by one set of vibrational numbers but having slightly different vibrational excitations belonging to the different spin states.

Transformation of an arbitrary density operator $\hat{\rho}(t)$ (NB!: being considered in the interaction representation) after the interaction with a microwave pulse of duration $\Delta t$ is given by
\begin{equation}
\hat\rho(t+\Delta t)\cong\hat{\boldsymbol{\cal U}}(\Delta t)\,\hat{\rho}(t)\,\hat{\boldsymbol{\cal U}}^{\dagger}(\Delta t),%
\label{a.12}%
\end{equation}
where $\hat{\rho}(t)$ can be an arbitrary either separable or entangled quantum state shared between the vibrational and spin subsystems of the atom. We can use this basic solution to manipulate qubits by a pulse sequence. Let us point out once again that the constructed solution only approximates the true dynamics of an atom driven by a microwave radiation, which becomes more valid as the difference between the trap potentials $U_a(\mathbf{r})$ and $U_b(\mathbf{r})$ reduces.

As an important example of application of (\ref{a.12}), consider the situation with Ramsey resonance and spin-echo protocol. Let the atom initially occupy the lower spin state $|a\rangle$ with a thermal distribution over its vibrational modes such that
\begin{equation}
\hat{\rho}(0)=\sum_{\mathbf{v}}\exp\left[\beta\left({\cal F}-\epsilon_{\mathbf{v}}\right)\right]|\mathbf{v};a\rangle\langle \mathbf{v};a|,%
\label{a.13}
\end{equation}
where $\beta$ is the inverse temperature and ${\cal F}={\cal F}(\beta)$ is its free energy. At the initial moment $t_{\pi/2} = 0$ the atom is excited by a short $\pi/2$-pulse. Then after the stage of free dynamics at a given time $t=t_{\pi} - t_{\pi/2}$ let the atom be excited by a short $\pi$-pulse, where $t_{\pi/2}, t_{\pi}$ are the arrival times of the pulses. As we see from the fundamental solution (\ref{a.8}), (\ref{a.10}) and (\ref{a.11}) such specific excitation transposes the probability amplitudes for any pairs of the $|\mathbf{v};a\rangle$ and $|\mathbf{w};b\rangle$ states with $\mathbf{w}=\mathbf{v}$. If after a certain delay $t = t_{\pi} - t_{\pi/2} = t'_{\pi/2} - t_{\pi}$ at the time $t'=2 t = t'_{\pi/2}$ the atom was excited by another $\pi/2$ pulse, it would come back to its initial state with the original density operator. In result of all the transformations at the time $t'=2 t$ we get $\hat{\rho}(t')=\hat{\rho}(0)$. It justifies that for an arbitrary number of spin qubits such a spin-echo protocol suggests preparation of the quantum register up to any given time $t'=2 t$ even in an array of relatively warm and mobile trapped atoms.

\section{Master equation for the density matrix}\label{Appendix_B}

\noindent Let us associate the environment with a fast recovering external reservoir unaffected by the atomic subsystem and uncoupled with it at any moment of time. Then the entire system has a density matrix  $\hat{\rho}_{\Sigma}=\hat{\rho}_{\Sigma}(t)=\hat{\rho}(t)\hat{\rho}_{R}$, i.e. its density operator factorizes to a product of the atomic $\hat{\rho}(t)$ and reservoir (environment) $\hat{\rho}_{R}$ components. The latter should be independent on time. Rigorously this assumption is not exactly valid and can be only approximately fulfilled and applicable for a smoothed behavior on a ``coarse grained'' time scale (i.e. for the increments longer than the correlation time of the driving Wiener-type stochastic process displaying the reservoir inertia).

The joint dynamics is described by the following quantum Liouville equation
\begin{eqnarray}
\frac{d\hat{\rho}_{\Sigma}}{d t}&=& -\frac{i}{\hbar} \left[\hat{H}_{\Sigma},\,\hat{\rho}_{\Sigma}(t)\right] -\frac{i}{\hbar} \left[\hat{W},\,\hat{\rho}_{\Sigma}(t)\right]%
\nonumber\\%
\nonumber\\%
\hat{H}_{\Sigma}&=&\hat{H}_{0}+\hat{H}_{R},
\label{b.1}
\end{eqnarray}
where the atomic Hamiltonian $\hat{H}_{0}$ is defined by (\ref{2.3}), $\hat{H}_{R}$ is the internal Hamiltonian of the reservoir, and the interaction term $\hat{W}$ is given by (\ref{2.12}). This equation is a precursor of (\ref{2.2}) specified for the extended density operator.

Let us transform (\ref{b.1}) to the interaction representation generated by the undisturbed Hamiltonian $\hat{H}_{\Sigma}$
\begin{eqnarray}
\tilde{W}(t)&=&\mathrm{e}^{\displaystyle\frac{i}{\hbar}\hat{H}_{\Sigma}t}\,\hat{W}\,\mathrm{e}^{-\displaystyle\frac{i}{\hbar}\hat{H}_{\Sigma}t},%
\nonumber\\%
\tilde{\rho}_{\Sigma}(t)&=&\mathrm{e}^{\displaystyle\frac{i}{\hbar}\hat{H}_{\Sigma}t}\,\hat{\rho}_{\Sigma}(t)\,\mathrm{e}^{-\displaystyle\frac{i}{\hbar}\hat{H}_{\Sigma}t},%
\nonumber\\%
\nonumber\\%
\dot{\tilde{\rho}}_{\Sigma}(t)&=&-\frac{i}{\hbar}\left[\tilde{W}(t),\tilde{\rho}_{\Sigma}(t)\right].%
\label{b.2}%
\end{eqnarray}
Then after subsequent iteration up to the second order and with partial tracing over the environment variables the last equation leads to the following increment of the atomic density operator:
\begin{widetext}
\begin{eqnarray}
\lefteqn{\tilde{\rho}(t+\Delta t)-\tilde{\rho}(t)\simeq\left(-\frac{i}{\hbar}\right)^2\!\int_t^{t+\Delta t}\!dt'\int_t^{t'}\!dt''}
\nonumber\\%
&&\times\left[\langle\tilde{W}(t')\tilde{W}(t'')\rangle\,\tilde{\rho}(t)-\langle\tilde{W}(t')\,\tilde{\rho}(t)\,\tilde{W}(t'')\rangle%
-\langle\tilde{W}(t'')\,\tilde{\rho}(t)\,\tilde{W}(t')\rangle+\tilde{\rho}(t)\langle\tilde{W}(t')\tilde{W}(t'')\rangle\right],%
\nonumber\\%
\label{b.3}%
\end{eqnarray}
\end{widetext}
where the angle brackets denote the partial trace $\langle\ldots\rangle\equiv\mathrm{Tr}'\tilde{\rho}_{R}\ldots$ and the contribution of the first order in the iteration process has vanished because of $\langle \tilde{W}(t')\rangle=0$. Here the time increment $\Delta t$ has to be much longer than the correlation time for the reservoir fluctuations but at the same time is sufficiently short to provide only a small increment for the atomic density matrix.

As was noted in section \ref{Subsection_II.D} there are two physical mechanisms initiating irreversible decoherence, which do not interfere with each other and can be separately described. Addressing to the master equation (\ref{2.2}) introduced in a symbolic form we can decompose the relaxation term into the sum of two contributions
\begin{equation}
\left( \frac{\partial\hat{\rho}}{\partial t}\right)_{\mathrm{rel}}=\left( \frac{\partial\hat{\rho}}{\partial t}\right)_{\mathrm{fl}}+\left( \frac{\partial\hat{\rho}}{\partial t}\right)_{\mathrm{sc}},%
\label{b.4}
\end{equation}
where in the right-hand side the first contribution is initiated by the intensity fluctuations of the laser light confining the atom, and the second term is the decoherence due to incoherent scattering of this light from the atom.

\subsubsection*{Intensity fluctuations}

\noindent The former process can be accurately modeled by classical stochastic fluctuations of the trap potential as follows:
\begin{eqnarray}
\hat{W}&\Rightarrow& \delta U_{0}(\hat{\mathbf{r}},t)=\xi(t)\,U_{0}(\hat{\mathbf{r}})%
\nonumber\\
\tilde{W}(t)&\Rightarrow&\xi(t)\,U_{0}(\hat{\mathbf{r}}(t)),%
\label{b.5}
\end{eqnarray}
where $\xi(t)$ is a classical Wiener-type stochastic process fluctuating near zero level with $\bar{\xi^2}\ll 1$. So we have assumed that only the total flux of the laser radiation can fluctuate preserving the potential shape.\footnote{We presume that the pointing noise of the trap position is negligible for our experimental setup, since, typically, for trap frequencies of tens of kHz the corresponding heating rate is much smaller than that due to intensity fluctuations \cite{LaserNoise1997}.}

In the steady state regime we define the correlation function
\begin{equation}
\langle\xi(t')\xi(t'')\rangle=g(t'-t''),%
\label{b.6}%
\end{equation}
where the angle brackets are now understood as a classical averaging. The spectral density of this stochastic process is given by
\begin{equation}
(\xi^2)_{\omega}=\int_{-\infty}^{\infty}d\tau\,g(\tau)\,\mathrm{e}^{i\omega t},%
\label{b.7}%
\end{equation}
which in our estimates has an infinitely broad spectrum in comparison with the trap oscillator frequencies contributing to (\ref{a.7}), so the function $g=g(\tau)$ has an infinitely short correlation time. We experimentally measured the intensity noise spectrum of the diode laser utilized in our dipole trap. The noise spectrum was observed in a spectral band between 0 and 500 MHz and shown to have a flat shape with relative intensity noise value estimated as $(\xi^2)_{\omega} \simeq 10^{-13} \, {\rm Hz}^{-1}$. Shot noise contribution is of order $10^{-16} \, {\rm Hz}^{-1}$ at optical power of several mW and is negligible.

Under the assumption of the zero-correlation time in (\ref{b.5}) and after evaluation of the time integrals in (\ref{b.3}) we arrive to the following contribution to (\ref{b.4})
\begin{eqnarray}
\lefteqn{\left( \frac{\partial\hat{\rho}}{\partial t}\right)_{\mathrm{fl}}=}%
\nonumber\\%
&&=-\frac{(\xi^2)_{0}}{2\hbar^2}\left[U_{0}^2(\hat{\mathbf{r}})\,\hat{\rho}(t)-2\,U_{0}(\hat{\mathbf{r}})\,\hat{\rho}(t)\,U_{0}(\hat{\mathbf{r}})+\hat{\rho}(t)U_{0}^2(\hat{\mathbf{r}})\right],%
\nonumber\\%
\label{b.8}
\end{eqnarray}
where, as was pointed above, the spectral density $(\xi^2)_{\omega=0}$ can be independently measured in experiment. Here, as in (\ref{2.2}) and (\ref{b.4}), we have assumed the original Schr\"{o}dinger representation.

\subsubsection*{Incoherent scattering}

\noindent For the process of incoherent scattering the perturbation Hamiltonian (\ref{2.12}) can be expressed by the scattering tensor of a two-level atom. Each integrand term contributes to (\ref{b.3}) in the interaction representation with exponential factors oscillating in time on the transition frequency for a particular scattering channel. The time integrals can be evaluated and the increment $\Delta t$ can be thinkable as long as to approximate the result by a $\delta$-function providing the energy conservation in the scattering process. Then each term in (\ref{b.3}) is treated in terms of the Fermi golden rule, reproducing the transition rates associated with either Rayleigh or Raman scattering events.

There are two types of contributions to the kinetic balance, namely out- and in-scattering terms. The out-scattering process either leaves (Rayleigh) or takes the atom away (Raman) from any state probabilistically occupied at a given time. Such process is parameterized by the total cross-section of light scattering, which without any relativistic corrections coincides with the scattering cross-section for a free atom
\begin{equation}
\sigma_0\simeq\sigma_0(\omega_L)=\frac{8\pi d_0^4}{3\hbar^2\!c^4}\frac{\omega_L^4}{(\omega_L-\omega_0)^2},%
\label{b.9}%
\end{equation}
where $d_0$ is the modulus of the dipole moment for the $S\to P$ optical transitions of such a non-relativistic alkali-metal atom. Within the applied approximations the total cross-section is independent on the vibrational number for any occupied state.

The in-scattering process is parameterized by the differential cross-section
\begin{eqnarray}
\lefteqn{d\sigma(\mathbf{k}_L,\mathbf{v},\mathbf{v}'\to \mathbf{k},\mathbf{v}'',\mathbf{v}''')\simeq}%
\nonumber\\%
&&\simeq d\sigma(\omega_L)\langle\mathbf{v}''|\mathrm{e}^{-i(\mathbf{k}-\mathbf{k}_L)\cdot\hat{\mathbf{r}}}|\mathbf{v}\rangle%
\langle\mathbf{v}'|\mathrm{e}^{i(\mathbf{k}-\mathbf{k}_L)\cdot\hat{\mathbf{r}}}|\mathbf{v}'''\rangle,%
\nonumber\\%
\label{b.10}%
\end{eqnarray}
which transfers the vibrational coherence, associated with dyad $|\mathbf{v}'\rangle\langle\mathbf{v}|$, to the coherence $|\mathbf{v}'''\rangle\langle\mathbf{v}''|$ as a result of the photon scattering from the laser mode $\mathbf{k}_L$ to the outgoing mode $\mathbf{k}$. The energy conservation dictates that $\epsilon_{\mathbf{v}''}-\epsilon_{\mathbf{v}}=\epsilon_{\mathbf{v}'''}-\epsilon_{\mathbf{v}'}$. The outer factor of the angular distribution is the differential cross-section for a free atom, which in the case of the incident light propagating along $z$-direction and linearly polarized along $x$-direction is given by
\begin{equation}
d\sigma(\omega_L)=\sigma_0(\omega_L)\frac{3}{8\pi}\left(1-\sin^2\theta\cos^2\phi\right)d\Omega_{\mathbf{k}},%
\label{b.11}%
\end{equation}
where $\{\theta,\phi\}=\Omega_{\mathbf{k}}$ is the solid angle for the scattered photon. The cross-section (\ref{b.10}) contributes to (\ref{b.3}) being integrated over all the scattering directions parameterized by the solid angle $\Omega_{\mathbf{k}}$. Then the coherence itself cannot be spontaneously induced in the system and only population transfer survives and contributes in the repopulation process initiated by the incoherent scattering of the laser light.

Eventually we arrive to the following contribution to the master equation:
\begin{eqnarray}
\lefteqn{\left( \frac{\partial\hat{\rho}}{\partial t}\right)_{\mathrm{sc}}=-\bar{I}\,\sigma_0\;\hat{\rho}(t)}%
\nonumber\\%
&&+\sum_{\mathbf{v},\mathbf{v}'}\sum_{\mathbf{v}'',\mathbf{v}'''}\,\int\! \bar{I}\,d\sigma(\mathbf{k}_L,\mathbf{v},\mathbf{v}'\to \mathbf{k},\mathbf{v}'',\mathbf{v}''')%
\nonumber\\%
&&\times  \sum_{m=a,b}\,\sum_{m'=a,b}|\mathbf{v}'';m\rangle\langle\mathbf{v};m|\ \hat{\rho}(t)\ |\mathbf{v}';m'\rangle\langle\mathbf{v}''';m'|,
\nonumber\\%
\label{b.12}
\end{eqnarray}
where $\bar{I}$ is the intensity of the photon flux (i.e. number of photons emitted per unit time and crossing a unit area) in the focal point of the laser beam.

Similarly to the laser fluctuations (\ref{b.8}) the process of incoherent scattering itself does not change the spin state of the atom. Again we obtain that the spin decoherence appears because of the difference in the oscillator energies (vibrational frequencies) for the $|a\rangle$ and $|b\rangle$ spin states. It is also noteworthy to point out that in a tight trap and in conditions of the strong Lamb-Dicke effect, when the exponent in (\ref{b.10}) could be approximated by a unit factor, the right hand side in (\ref{b.12}) vanishes. In other words, the elastic Rayleigh scattering does not disturb coherence in the spin dynamics, which was earlier pointed in \cite{Meschede2005}.

Now the master equation (\ref{2.2}) is presented in a closed form and we can clarify the solution procedure, which we shall further follow. We will keep in mind the following two basic transformations of the density operator. One is the qubit control by a short microwave pulse. During this excitation there is no decoherence and we apply the transformation (\ref{a.12}) to a currently existing value of the density operator. In the intermediate stages between the rounds of the microwave chirping we solve the Cauchy problem for the master equation driven by the system Hamiltonian and by interaction with the environment.

For a solution we define the following slow-varying components of the density matrix
\begin{eqnarray}%
\tilde{\rho}_{b,\mathbf{w};a,\mathbf{v}}(t)&=&\exp\left[+i\left(\omega_{\mathrm{hpf}}+\omega_{\mathbf{w}\mathbf{v}}\right)t\right]\,\rho_{b,\mathbf{w};a,\mathbf{v}}(t)%
\nonumber\\%
\tilde{\rho}_{a,\mathbf{v};b,\mathbf{w}}(t)&=&\exp\left[+i\left(-\omega_{\mathrm{hpf}}+\omega_{\mathbf{v}\mathbf{w}}\right)t\right]\,\rho_{a,\mathbf{v};b,\mathbf{w}}(t)%
\nonumber\\%
\tilde{\rho}_{a,\mathbf{v'};a,\mathbf{v}}(t)&=&\exp\left[+i\,\omega_{\mathbf{v}',\mathbf{v}}t\right]\,\rho_{a,\mathbf{v'};a,\mathbf{v}}(t)%
\nonumber\\%
\tilde{\rho}_{b,\mathbf{w}';b,\mathbf{w}}(t)&=&\exp\left[+i\,\omega_{\mathbf{w}',\mathbf{w}}t\right]\,\rho_{b,\mathbf{w}';b,\mathbf{w}}(t)%
\label{b.13},%
\end{eqnarray}%
and substitute them into the master equation in its basic form (\ref{2.2}).

Here again we have distinguished the vibrational numbers for the lower and upper hyperfine sublevels as the energy difference is critically important for the description of a free precession of the spin-vibrational coherences. Since the decoherence is expected to be a long developing process, we can take into concideration that the transverse vibrational modes are actually non-degenerate with slightly different eigenfrequencies, which eliminates the generation of vibrational coherency from the original thermal state (\ref{a.13}) via both the dissipative mechanisms either (\ref{b.8}) or (\ref{b.12}). Thus with keeping only diagonal elements of the density matrix with respect to the vibrational numbers with $\mathbf{v}=\mathbf{w}$ we obtain
\begin{eqnarray}
\dot{\tilde{\rho}}_{b,\mathbf{v};a,\mathbf{v}}(t)&=& -\Gamma_{\mathbf{v}}\tilde{\rho}_{b,\mathbf{v};a,\mathbf{v}}(t)%
\nonumber\\%
&&+{\sum_{\mathbf{v}'\neq\mathbf{v}}}^{\prime}\mathrm{e}^{+i\delta\omega_{\mathbf{v}\mathbf{v}'}t}\,%
\Gamma_{\mathbf{v}\mathbf{v}'}\,\tilde{\rho}_{b,\mathbf{v}';a,\mathbf{v}'}(t)\,%
\nonumber\\%
\dot{\tilde{\rho}}_{a,\mathbf{v};b,\mathbf{v}}(t)&=&  -\Gamma_{\mathbf{v}}\tilde{\rho}_{a,\mathbf{v};b,\mathbf{v}}(t)%
\nonumber\\%
&&+{\sum_{\mathbf{v}'\neq\mathbf{v}}}^{\prime}\mathrm{e}^{-i\delta\omega_{\mathbf{v}\mathbf{v}'}t}\,%
\Gamma_{\mathbf{v}\mathbf{v}'}\,\tilde{\rho}_{a,\mathbf{v}';b,\mathbf{v}'}(t)\,%
\nonumber\\%
\dot{\tilde{\rho}}_{a,\mathbf{v};a,\mathbf{v}}(t)&=&  -\Gamma_{\mathbf{v}}\tilde{\rho}_{a,\mathbf{v};a,\mathbf{v}}(t)%
+{\sum_{\mathbf{v}'\neq\mathbf{v}}}^{\prime}\,%
\Gamma_{\mathbf{v}\mathbf{v}'}\,\tilde{\rho}_{a,\mathbf{v}';a,\mathbf{v}'}(t)\,%
\nonumber\\%
\dot{\tilde{\rho}}_{b,\mathbf{v};b,\mathbf{v}}(t)&=&  -\Gamma_{\mathbf{v}}\tilde{\rho}_{b,\mathbf{v};b,\mathbf{v}}(t)%
+{\sum_{\mathbf{v}'\neq\mathbf{v}}}^{\prime}\,%
\Gamma_{\mathbf{v}\mathbf{v}'}\,\tilde{\rho}_{b,\mathbf{v}';b,\mathbf{v}'}(t),\,%
\nonumber\\%
\label{b.14}%
\end{eqnarray}
where
\begin{equation}
\delta\omega_{\mathbf{v}\mathbf{v}'}=\delta\omega_{\mathbf{v}}-\delta\omega_{\mathbf{v}'}%
\label{b.15}
\end{equation}
and $\delta\omega_{\mathbf{v}}$ is defined by (\ref{a.7}).

In accordance with (\ref{b.4}) the relaxation matrix consists of two contributions
\begin{eqnarray}
\Gamma_{\mathbf{v}}&=&\Gamma_{\mathbf{v}}^{(\mathrm{fl})}+\Gamma_{\mathbf{v}}^{(\mathrm{sc})}%
\nonumber\\%
\Gamma_{\mathbf{v}\mathbf{v}'}&=&\Gamma_{\mathbf{v}\mathbf{v}'}^{(\mathrm{fl})}+\Gamma_{\mathbf{v}\mathbf{v}'}^{(\mathrm{sc})},%
\label{b.16}%
\end{eqnarray}
where
\begin{eqnarray}
\Gamma_{\mathbf{v}}^{(\mathrm{fl})}&=&\frac{(\xi^2)_{0}}{\hbar^2}\,\left[\langle\mathbf{v}|U_{0}^2(\hat{\mathbf{r}})|\mathbf{v}\rangle%
-\left|\langle\mathbf{v}|U_{0}(\hat{\mathbf{r}})|\mathbf{v}\rangle\right|^2\right]%
\nonumber\\%
\Gamma_{\mathbf{v}\mathbf{v}'}^{(\mathrm{fl})}&=&\frac{(\xi^2)_{0}}{\hbar^2}\,\left|\langle\mathbf{v}|U_{0}(\hat{\mathbf{r}})|\mathbf{v}'\rangle\right|^2
\label{b.17}%
\end{eqnarray}
and
\begin{eqnarray}
\Gamma_{\mathbf{v}}^{(\mathrm{sc})}&=&\bar{I}\,\sigma_0\left[1-\left\langle\left|\langle\mathbf{v}|\mathrm{e}^{-i(\mathbf{k}-\mathbf{k}_L)\cdot\hat{\mathbf{r}}}|\mathbf{v}\rangle\right|^{2}\right\rangle\right]%
\nonumber\\%
\Gamma_{\mathbf{v}\mathbf{v}'}^{(\mathrm{sc})}&=&\bar{I}\,\sigma_0\left\langle\left|\langle\mathbf{v}|\mathrm{e}^{-i(\mathbf{k}-\mathbf{k}_L)\cdot\hat{\mathbf{r}}}|\mathbf{v}'\rangle\right|^{2}\right\rangle,%
\label{b.18}%
\end{eqnarray}
where outer angle brackets denote the angle averaging over the scattering directions in accordance with (\ref{b.11}).

As one can see, equations (\ref{b.14}) evidently fulfil the following steady state solution
\begin{eqnarray}
\tilde{\rho}_{b,\mathbf{v};a,\mathbf{v}}(\infty)&=&\tilde{\rho}_{a,\mathbf{v};b,\mathbf{v}}(\infty)=0
\nonumber\\%
\tilde{\rho}_{a,\mathbf{v};a,\mathbf{v}}(\infty)&=&\mathrm{const}_a%
\nonumber\\%
\tilde{\rho}_{b,\mathbf{v};b,\mathbf{v}}(\infty)&=&\mathrm{const}_b,%
\label{b.19}%
\end{eqnarray}
which constitutes that the laser noise, converted into the parametric heating process, as well as the incoherent Raman emission increase the atom's temperature up to infinite level. But both the processes develop in such a way that the repopulation of the vibrational states does not influence the occupation probabilities of the spin states. It is also evident that in an ideal case if $\delta\omega_{\mathbf{v}\mathbf{v}'}\to 0$ the equations (\ref{b.14}) can be traced over the vibrational degrees of freedom and both the processes would not affect the spin dynamics at all. This option can be approximately adjusted with extra cooling of the atom down to the ground state of the oscillator well. For such an optimal scenario it becomes important to take into account the relativistic corrections to the scattering process, which would extend the coherence times up to seconds, as shown in Fig.{\ref{fig4}} \cite{Heinzen1994,Meshede2000}.

\bibliographystyle{apsrev4-1}
\bibliography{references}

\begin{thebibliography}{34}%
\makeatletter
\providecommand \@ifxundefined [1]{%
 \@ifx{#1\undefined}
}%
\providecommand \@ifnum [1]{%
 \ifnum #1\expandafter \@firstoftwo
 \else \expandafter \@secondoftwo
 \fi
}%
\providecommand \@ifx [1]{%
 \ifx #1\expandafter \@firstoftwo
 \else \expandafter \@secondoftwo
 \fi
}%
\providecommand \natexlab [1]{#1}%
\providecommand \enquote  [1]{``#1''}%
\providecommand \bibnamefont  [1]{#1}%
\providecommand \bibfnamefont [1]{#1}%
\providecommand \citenamefont [1]{#1}%
\providecommand \href@noop [0]{\@secondoftwo}%
\providecommand \href [0]{\begingroup \@sanitize@url \@href}%
\providecommand \@href[1]{\@@startlink{#1}\@@href}%
\providecommand \@@href[1]{\endgroup#1\@@endlink}%
\providecommand \@sanitize@url [0]{\catcode `\\12\catcode `\$12\catcode
  `\&12\catcode `\#12\catcode `\^12\catcode `\_12\catcode `\%12\relax}%
\providecommand \@@startlink[1]{}%
\providecommand \@@endlink[0]{}%
\providecommand \url  [0]{\begingroup\@sanitize@url \@url }%
\providecommand \@url [1]{\endgroup\@href {#1}{\urlprefix }}%
\providecommand \urlprefix  [0]{URL }%
\providecommand \Eprint [0]{\href }%
\providecommand \doibase [0]{http://dx.doi.org/}%
\providecommand \selectlanguage [0]{\@gobble}%
\providecommand \bibinfo  [0]{\@secondoftwo}%
\providecommand \bibfield  [0]{\@secondoftwo}%
\providecommand \translation [1]{[#1]}%
\providecommand \BibitemOpen [0]{}%
\providecommand \bibitemStop [0]{}%
\providecommand \bibitemNoStop [0]{.\EOS\space}%
\providecommand \EOS [0]{\spacefactor3000\relax}%
\providecommand \BibitemShut  [1]{\csname bibitem#1\endcsname}%
\let\auto@bib@innerbib\@empty
\bibitem [{\citenamefont {Graham}\ \emph {et~al.}(2019)\citenamefont {Graham},
  \citenamefont {Kwon}, \citenamefont {Grinkemeyer}, \citenamefont {Marra},
  \citenamefont {Jiang}, \citenamefont {Lichtman}, \citenamefont {Sun},
  \citenamefont {Ebert},\ and\ \citenamefont {Saffman}}]{Saffman2019}%
  \BibitemOpen
  \bibfield  {author} {\bibinfo {author} {\bibfnamefont {T.~M.}\ \bibnamefont
  {Graham}}, \bibinfo {author} {\bibfnamefont {M.}~\bibnamefont {Kwon}},
  \bibinfo {author} {\bibfnamefont {B.}~\bibnamefont {Grinkemeyer}}, \bibinfo
  {author} {\bibfnamefont {Z.}~\bibnamefont {Marra}}, \bibinfo {author}
  {\bibfnamefont {X.}~\bibnamefont {Jiang}}, \bibinfo {author} {\bibfnamefont
  {M.~T.}\ \bibnamefont {Lichtman}}, \bibinfo {author} {\bibfnamefont
  {Y.}~\bibnamefont {Sun}}, \bibinfo {author} {\bibfnamefont {M.}~\bibnamefont
  {Ebert}}, \ and\ \bibinfo {author} {\bibfnamefont {M.}~\bibnamefont
  {Saffman}},\ }\href {\doibase 10.1103/PhysRevLett.123.230501} {\bibfield
  {journal} {\bibinfo  {journal} {Phys. Rev. Lett.}\ }\textbf {\bibinfo
  {volume} {123}},\ \bibinfo {pages} {230501} (\bibinfo {year}
  {2019})}\BibitemShut {NoStop}%
\bibitem [{\citenamefont {Levine}\ \emph {et~al.}(2019)\citenamefont {Levine},
  \citenamefont {Keesling}, \citenamefont {Semeghini}, \citenamefont {Omran},
  \citenamefont {Wang}, \citenamefont {Ebadi}, \citenamefont {Bernien},
  \citenamefont {Greiner}, \citenamefont {Vuleti\ifmmode~\acute{c}\else
  \'{c}\fi{}}, \citenamefont {Pichler},\ and\ \citenamefont
  {Lukin}}]{Lukin2019}%
  \BibitemOpen
  \bibfield  {author} {\bibinfo {author} {\bibfnamefont {H.}~\bibnamefont
  {Levine}}, \bibinfo {author} {\bibfnamefont {A.}~\bibnamefont {Keesling}},
  \bibinfo {author} {\bibfnamefont {G.}~\bibnamefont {Semeghini}}, \bibinfo
  {author} {\bibfnamefont {A.}~\bibnamefont {Omran}}, \bibinfo {author}
  {\bibfnamefont {T.~T.}\ \bibnamefont {Wang}}, \bibinfo {author}
  {\bibfnamefont {S.}~\bibnamefont {Ebadi}}, \bibinfo {author} {\bibfnamefont
  {H.}~\bibnamefont {Bernien}}, \bibinfo {author} {\bibfnamefont
  {M.}~\bibnamefont {Greiner}}, \bibinfo {author} {\bibfnamefont
  {V.}~\bibnamefont {Vuleti\ifmmode~\acute{c}\else \'{c}\fi{}}}, \bibinfo
  {author} {\bibfnamefont {H.}~\bibnamefont {Pichler}}, \ and\ \bibinfo
  {author} {\bibfnamefont {M.~D.}\ \bibnamefont {Lukin}},\ }\href {\doibase
  10.1103/PhysRevLett.123.170503} {\bibfield  {journal} {\bibinfo  {journal}
  {Phys. Rev. Lett.}\ }\textbf {\bibinfo {volume} {123}},\ \bibinfo {pages}
  {170503} (\bibinfo {year} {2019})}\BibitemShut {NoStop}%
\bibitem [{\citenamefont {Labuhn}\ \emph {et~al.}(2016)\citenamefont {Labuhn},
  \citenamefont {Barredo}, \citenamefont {Ravets}, \citenamefont
  {De~L{\'e}s{\'e}leuc}, \citenamefont {Macr{\`\i}}, \citenamefont {Lahaye},\
  and\ \citenamefont {Browaeys}}]{Browaeys_Nature2016}%
  \BibitemOpen
  \bibfield  {author} {\bibinfo {author} {\bibfnamefont {H.}~\bibnamefont
  {Labuhn}}, \bibinfo {author} {\bibfnamefont {D.}~\bibnamefont {Barredo}},
  \bibinfo {author} {\bibfnamefont {S.}~\bibnamefont {Ravets}}, \bibinfo
  {author} {\bibfnamefont {S.}~\bibnamefont {De~L{\'e}s{\'e}leuc}}, \bibinfo
  {author} {\bibfnamefont {T.}~\bibnamefont {Macr{\`\i}}}, \bibinfo {author}
  {\bibfnamefont {T.}~\bibnamefont {Lahaye}}, \ and\ \bibinfo {author}
  {\bibfnamefont {A.}~\bibnamefont {Browaeys}},\ }\href@noop {} {\bibfield
  {journal} {\bibinfo  {journal} {Nature}\ }\textbf {\bibinfo {volume} {534}},\
  \bibinfo {pages} {667} (\bibinfo {year} {2016})}\BibitemShut {NoStop}%
\bibitem [{\citenamefont {Keesling}\ \emph {et~al.}(2019)\citenamefont
  {Keesling}, \citenamefont {Omran}, \citenamefont {Levine}, \citenamefont
  {Bernien}, \citenamefont {Pichler}, \citenamefont {Choi}, \citenamefont
  {Samajdar}, \citenamefont {Schwartz}, \citenamefont {Silvi}, \citenamefont
  {Sachdev} \emph {et~al.}}]{Lukin_Nature2019}%
  \BibitemOpen
  \bibfield  {author} {\bibinfo {author} {\bibfnamefont {A.}~\bibnamefont
  {Keesling}}, \bibinfo {author} {\bibfnamefont {A.}~\bibnamefont {Omran}},
  \bibinfo {author} {\bibfnamefont {H.}~\bibnamefont {Levine}}, \bibinfo
  {author} {\bibfnamefont {H.}~\bibnamefont {Bernien}}, \bibinfo {author}
  {\bibfnamefont {H.}~\bibnamefont {Pichler}}, \bibinfo {author} {\bibfnamefont
  {S.}~\bibnamefont {Choi}}, \bibinfo {author} {\bibfnamefont {R.}~\bibnamefont
  {Samajdar}}, \bibinfo {author} {\bibfnamefont {S.}~\bibnamefont {Schwartz}},
  \bibinfo {author} {\bibfnamefont {P.}~\bibnamefont {Silvi}}, \bibinfo
  {author} {\bibfnamefont {S.}~\bibnamefont {Sachdev}},  \emph {et~al.},\
  }\href@noop {} {\bibfield  {journal} {\bibinfo  {journal} {Nature}\ }\textbf
  {\bibinfo {volume} {568}},\ \bibinfo {pages} {207} (\bibinfo {year}
  {2019})}\BibitemShut {NoStop}%
\bibitem [{\citenamefont {Browaeys}\ and\ \citenamefont
  {Lahaye}(2020)}]{Browaeys_NaturePhys2020}%
  \BibitemOpen
  \bibfield  {author} {\bibinfo {author} {\bibfnamefont {A.}~\bibnamefont
  {Browaeys}}\ and\ \bibinfo {author} {\bibfnamefont {T.}~\bibnamefont
  {Lahaye}},\ }\href {\doibase 10.1038/s41567-019-0733-z} {\bibfield  {journal}
  {\bibinfo  {journal} {Nature Physics}\ }\textbf {\bibinfo {volume} {16}},\
  \bibinfo {pages} {132} (\bibinfo {year} {2020})}\BibitemShut {NoStop}%
\bibitem [{\citenamefont {Zurek}(1982)}]{Zurek1982}%
  \BibitemOpen
  \bibfield  {author} {\bibinfo {author} {\bibfnamefont {W.~H.}\ \bibnamefont
  {Zurek}},\ }\href {\doibase 10.1103/PhysRevD.26.1862} {\bibfield  {journal}
  {\bibinfo  {journal} {Phys. Rev. D}\ }\textbf {\bibinfo {volume} {26}},\
  \bibinfo {pages} {1862} (\bibinfo {year} {1982})}\BibitemShut {NoStop}%
\bibitem [{\citenamefont {Cline}\ \emph {et~al.}(1994)\citenamefont {Cline},
  \citenamefont {Miller}, \citenamefont {Matthews},\ and\ \citenamefont
  {Heinzen}}]{Heinzen1994}%
  \BibitemOpen
  \bibfield  {author} {\bibinfo {author} {\bibfnamefont {R.~A.}\ \bibnamefont
  {Cline}}, \bibinfo {author} {\bibfnamefont {J.~D.}\ \bibnamefont {Miller}},
  \bibinfo {author} {\bibfnamefont {M.~R.}\ \bibnamefont {Matthews}}, \ and\
  \bibinfo {author} {\bibfnamefont {D.~J.}\ \bibnamefont {Heinzen}},\ }\href
  {\doibase 10.1364/OL.19.000207} {\bibfield  {journal} {\bibinfo  {journal}
  {Opt. Lett.}\ }\textbf {\bibinfo {volume} {19}},\ \bibinfo {pages} {207}
  (\bibinfo {year} {1994})}\BibitemShut {NoStop}%
\bibitem [{\citenamefont {Grimm}\ \emph {et~al.}(2000)\citenamefont {Grimm},
  \citenamefont {Weidemüller},\ and\ \citenamefont {Ovchinnikov}}]{Grimm2000}%
  \BibitemOpen
  \bibfield  {author} {\bibinfo {author} {\bibfnamefont {R.}~\bibnamefont
  {Grimm}}, \bibinfo {author} {\bibfnamefont {M.}~\bibnamefont {Weidemüller}},
  \ and\ \bibinfo {author} {\bibfnamefont {Y.~B.}\ \bibnamefont {Ovchinnikov}}\
  }(\bibinfo  {publisher} {Academic Press},\ \bibinfo {year} {2000})\ pp.\
  \bibinfo {pages} {95 -- 170}\BibitemShut {NoStop}%
\bibitem [{\citenamefont {Savard}\ \emph
  {et~al.}(1997{\natexlab{a}})\citenamefont {Savard}, \citenamefont {O'Hara},\
  and\ \citenamefont {Thomas}}]{LaserNoise1997}%
  \BibitemOpen
  \bibfield  {author} {\bibinfo {author} {\bibfnamefont {T.~A.}\ \bibnamefont
  {Savard}}, \bibinfo {author} {\bibfnamefont {K.~M.}\ \bibnamefont {O'Hara}},
  \ and\ \bibinfo {author} {\bibfnamefont {J.~E.}\ \bibnamefont {Thomas}},\
  }\href {\doibase 10.1103/PhysRevA.56.R1095} {\bibfield  {journal} {\bibinfo
  {journal} {Phys. Rev. A}\ }\textbf {\bibinfo {volume} {56}},\ \bibinfo
  {pages} {R1095} (\bibinfo {year} {1997}{\natexlab{a}})}\BibitemShut {NoStop}%
\bibitem [{\citenamefont {Gardiner}\ \emph {et~al.}(2000)\citenamefont
  {Gardiner}, \citenamefont {Ye}, \citenamefont {Nagerl},\ and\ \citenamefont
  {Kimble}}]{Kimble2000}%
  \BibitemOpen
  \bibfield  {author} {\bibinfo {author} {\bibfnamefont {C.~W.}\ \bibnamefont
  {Gardiner}}, \bibinfo {author} {\bibfnamefont {J.}~\bibnamefont {Ye}},
  \bibinfo {author} {\bibfnamefont {H.~C.}\ \bibnamefont {Nagerl}}, \ and\
  \bibinfo {author} {\bibfnamefont {H.~J.}\ \bibnamefont {Kimble}},\ }\href
  {\doibase 10.1103/PhysRevA.61.045801} {\bibfield  {journal} {\bibinfo
  {journal} {Phys. Rev. A}\ }\textbf {\bibinfo {volume} {61}},\ \bibinfo
  {pages} {045801} (\bibinfo {year} {2000})}\BibitemShut {NoStop}%
\bibitem [{\citenamefont {Kuhr}\ \emph {et~al.}(2005)\citenamefont {Kuhr},
  \citenamefont {Alt}, \citenamefont {Schrader}, \citenamefont {Dotsenko},
  \citenamefont {Miroshnychenko}, \citenamefont {Rauschenbeutel},\ and\
  \citenamefont {Meschede}}]{Meschede2005}%
  \BibitemOpen
  \bibfield  {author} {\bibinfo {author} {\bibfnamefont {S.}~\bibnamefont
  {Kuhr}}, \bibinfo {author} {\bibfnamefont {W.}~\bibnamefont {Alt}}, \bibinfo
  {author} {\bibfnamefont {D.}~\bibnamefont {Schrader}}, \bibinfo {author}
  {\bibfnamefont {I.}~\bibnamefont {Dotsenko}}, \bibinfo {author}
  {\bibfnamefont {Y.}~\bibnamefont {Miroshnychenko}}, \bibinfo {author}
  {\bibfnamefont {A.}~\bibnamefont {Rauschenbeutel}}, \ and\ \bibinfo {author}
  {\bibfnamefont {D.}~\bibnamefont {Meschede}},\ }\href {\doibase
  10.1103/PhysRevA.72.023406} {\bibfield  {journal} {\bibinfo  {journal} {Phys.
  Rev. A}\ }\textbf {\bibinfo {volume} {72}},\ \bibinfo {pages} {023406}
  (\bibinfo {year} {2005})}\BibitemShut {NoStop}%
\bibitem [{\citenamefont {Windpassinger}\ \emph {et~al.}(2008)\citenamefont
  {Windpassinger}, \citenamefont {Oblak}, \citenamefont {Hoff}, \citenamefont
  {Appel}, \citenamefont {Kj{\ae}rgaard},\ and\ \citenamefont
  {Polzik}}]{Polzik2008}%
  \BibitemOpen
  \bibfield  {author} {\bibinfo {author} {\bibfnamefont {P.~J.}\ \bibnamefont
  {Windpassinger}}, \bibinfo {author} {\bibfnamefont {D.}~\bibnamefont
  {Oblak}}, \bibinfo {author} {\bibfnamefont {U.~B.}\ \bibnamefont {Hoff}},
  \bibinfo {author} {\bibfnamefont {J.}~\bibnamefont {Appel}}, \bibinfo
  {author} {\bibfnamefont {N.}~\bibnamefont {Kj{\ae}rgaard}}, \ and\ \bibinfo
  {author} {\bibfnamefont {E.~S.}\ \bibnamefont {Polzik}},\ }\href {\doibase
  10.1088/1367-2630/10/5/053032} {\bibfield  {journal} {\bibinfo  {journal}
  {New Journal of Physics}\ }\textbf {\bibinfo {volume} {10}},\ \bibinfo
  {pages} {053032} (\bibinfo {year} {2008})}\BibitemShut {NoStop}%
\bibitem [{\citenamefont {Gerbier}\ \emph {et~al.}(2006)\citenamefont
  {Gerbier}, \citenamefont {Widera}, \citenamefont {F\"olling}, \citenamefont
  {Mandel},\ and\ \citenamefont {Bloch}}]{Bloch2006}%
  \BibitemOpen
  \bibfield  {author} {\bibinfo {author} {\bibfnamefont {F.}~\bibnamefont
  {Gerbier}}, \bibinfo {author} {\bibfnamefont {A.}~\bibnamefont {Widera}},
  \bibinfo {author} {\bibfnamefont {S.}~\bibnamefont {F\"olling}}, \bibinfo
  {author} {\bibfnamefont {O.}~\bibnamefont {Mandel}}, \ and\ \bibinfo {author}
  {\bibfnamefont {I.}~\bibnamefont {Bloch}},\ }\href {\doibase
  10.1103/PhysRevA.73.041602} {\bibfield  {journal} {\bibinfo  {journal} {Phys.
  Rev. A}\ }\textbf {\bibinfo {volume} {73}},\ \bibinfo {pages} {041602}
  (\bibinfo {year} {2006})}\BibitemShut {NoStop}%
\bibitem [{\citenamefont {Jones}\ \emph {et~al.}(2007)\citenamefont {Jones},
  \citenamefont {Beugnon}, \citenamefont {Ga\"etan}, \citenamefont {Zhang},
  \citenamefont {Messin}, \citenamefont {Browaeys},\ and\ \citenamefont
  {Grangier}}]{Grangier2007}%
  \BibitemOpen
  \bibfield  {author} {\bibinfo {author} {\bibfnamefont {M.~P.~A.}\
  \bibnamefont {Jones}}, \bibinfo {author} {\bibfnamefont {J.}~\bibnamefont
  {Beugnon}}, \bibinfo {author} {\bibfnamefont {A.}~\bibnamefont {Ga\"etan}},
  \bibinfo {author} {\bibfnamefont {J.}~\bibnamefont {Zhang}}, \bibinfo
  {author} {\bibfnamefont {G.}~\bibnamefont {Messin}}, \bibinfo {author}
  {\bibfnamefont {A.}~\bibnamefont {Browaeys}}, \ and\ \bibinfo {author}
  {\bibfnamefont {P.}~\bibnamefont {Grangier}},\ }\href {\doibase
  10.1103/PhysRevA.75.040301} {\bibfield  {journal} {\bibinfo  {journal} {Phys.
  Rev. A}\ }\textbf {\bibinfo {volume} {75}},\ \bibinfo {pages} {040301}
  (\bibinfo {year} {2007})}\BibitemShut {NoStop}%
\bibitem [{\citenamefont {Karski}\ \emph {et~al.}(2010)\citenamefont {Karski},
  \citenamefont {Förster}, \citenamefont {Choi}, \citenamefont {Steffen},
  \citenamefont {Belmechri}, \citenamefont {Alt}, \citenamefont {Meschede},\
  and\ \citenamefont {Widera}}]{Meshede2010}%
  \BibitemOpen
  \bibfield  {author} {\bibinfo {author} {\bibfnamefont {M.}~\bibnamefont
  {Karski}}, \bibinfo {author} {\bibfnamefont {L.}~\bibnamefont {Förster}},
  \bibinfo {author} {\bibfnamefont {J.-M.}\ \bibnamefont {Choi}}, \bibinfo
  {author} {\bibfnamefont {A.}~\bibnamefont {Steffen}}, \bibinfo {author}
  {\bibfnamefont {N.}~\bibnamefont {Belmechri}}, \bibinfo {author}
  {\bibfnamefont {W.}~\bibnamefont {Alt}}, \bibinfo {author} {\bibfnamefont
  {D.}~\bibnamefont {Meschede}}, \ and\ \bibinfo {author} {\bibfnamefont
  {A.}~\bibnamefont {Widera}},\ }\href {\doibase 10.1088/1367-2630/12/6/065027}
  {\bibfield  {journal} {\bibinfo  {journal} {New Journal of Physics}\ }\textbf
  {\bibinfo {volume} {12}},\ \bibinfo {pages} {065027} (\bibinfo {year}
  {2010})}\BibitemShut {NoStop}%
\bibitem [{\citenamefont {Xia}\ \emph {et~al.}(2015)\citenamefont {Xia},
  \citenamefont {Lichtman}, \citenamefont {Maller}, \citenamefont {Carr},
  \citenamefont {Piotrowicz}, \citenamefont {Isenhower},\ and\ \citenamefont
  {Saffman}}]{Saffman2015}%
  \BibitemOpen
  \bibfield  {author} {\bibinfo {author} {\bibfnamefont {T.}~\bibnamefont
  {Xia}}, \bibinfo {author} {\bibfnamefont {M.}~\bibnamefont {Lichtman}},
  \bibinfo {author} {\bibfnamefont {K.}~\bibnamefont {Maller}}, \bibinfo
  {author} {\bibfnamefont {A.~W.}\ \bibnamefont {Carr}}, \bibinfo {author}
  {\bibfnamefont {M.~J.}\ \bibnamefont {Piotrowicz}}, \bibinfo {author}
  {\bibfnamefont {L.}~\bibnamefont {Isenhower}}, \ and\ \bibinfo {author}
  {\bibfnamefont {M.}~\bibnamefont {Saffman}},\ }\href {\doibase
  10.1103/PhysRevLett.114.100503} {\bibfield  {journal} {\bibinfo  {journal}
  {Phys. Rev. Lett.}\ }\textbf {\bibinfo {volume} {114}},\ \bibinfo {pages}
  {100503} (\bibinfo {year} {2015})}\BibitemShut {NoStop}%
\bibitem [{\citenamefont {Wang}\ \emph {et~al.}(2016)\citenamefont {Wang},
  \citenamefont {Kumar}, \citenamefont {Wu},\ and\ \citenamefont
  {Weiss}}]{Weiss2016}%
  \BibitemOpen
  \bibfield  {author} {\bibinfo {author} {\bibfnamefont {Y.}~\bibnamefont
  {Wang}}, \bibinfo {author} {\bibfnamefont {A.}~\bibnamefont {Kumar}},
  \bibinfo {author} {\bibfnamefont {T.-Y.}\ \bibnamefont {Wu}}, \ and\ \bibinfo
  {author} {\bibfnamefont {D.~S.}\ \bibnamefont {Weiss}},\ }\href {\doibase
  10.1126/science.aaf2581} {\bibfield  {journal} {\bibinfo  {journal}
  {Science}\ }\textbf {\bibinfo {volume} {352}},\ \bibinfo {pages} {1562}
  (\bibinfo {year} {2016})}\BibitemShut {NoStop}%
\bibitem [{\citenamefont {Sheng}\ \emph
  {et~al.}(2018{\natexlab{a}})\citenamefont {Sheng}, \citenamefont {He},
  \citenamefont {Xu}, \citenamefont {Guo}, \citenamefont {Wang}, \citenamefont
  {Xiong}, \citenamefont {Liu}, \citenamefont {Wang},\ and\ \citenamefont
  {Zhan}}]{Mingsheng2018}%
  \BibitemOpen
  \bibfield  {author} {\bibinfo {author} {\bibfnamefont {C.}~\bibnamefont
  {Sheng}}, \bibinfo {author} {\bibfnamefont {X.}~\bibnamefont {He}}, \bibinfo
  {author} {\bibfnamefont {P.}~\bibnamefont {Xu}}, \bibinfo {author}
  {\bibfnamefont {R.}~\bibnamefont {Guo}}, \bibinfo {author} {\bibfnamefont
  {K.}~\bibnamefont {Wang}}, \bibinfo {author} {\bibfnamefont {Z.}~\bibnamefont
  {Xiong}}, \bibinfo {author} {\bibfnamefont {M.}~\bibnamefont {Liu}}, \bibinfo
  {author} {\bibfnamefont {J.}~\bibnamefont {Wang}}, \ and\ \bibinfo {author}
  {\bibfnamefont {M.}~\bibnamefont {Zhan}},\ }\href {\doibase
  10.1103/PhysRevLett.121.240501} {\bibfield  {journal} {\bibinfo  {journal}
  {Phys. Rev. Lett.}\ }\textbf {\bibinfo {volume} {121}},\ \bibinfo {pages}
  {240501} (\bibinfo {year} {2018}{\natexlab{a}})}\BibitemShut {NoStop}%
\bibitem [{\citenamefont {Derevianko}(2010)}]{Derevianko_PRA2010}%
  \BibitemOpen
  \bibfield  {author} {\bibinfo {author} {\bibfnamefont {A.}~\bibnamefont
  {Derevianko}},\ }\href {\doibase 10.1103/PhysRevA.81.051606} {\bibfield
  {journal} {\bibinfo  {journal} {Phys. Rev. A}\ }\textbf {\bibinfo {volume}
  {81}},\ \bibinfo {pages} {051606} (\bibinfo {year} {2010})}\BibitemShut
  {NoStop}%
\bibitem [{\citenamefont {Kaufman}\ \emph {et~al.}(2012)\citenamefont
  {Kaufman}, \citenamefont {Lester},\ and\ \citenamefont
  {Regal}}]{Regal_PRX2012}%
  \BibitemOpen
  \bibfield  {author} {\bibinfo {author} {\bibfnamefont {A.~M.}\ \bibnamefont
  {Kaufman}}, \bibinfo {author} {\bibfnamefont {B.~J.}\ \bibnamefont {Lester}},
  \ and\ \bibinfo {author} {\bibfnamefont {C.~A.}\ \bibnamefont {Regal}},\
  }\href {\doibase 10.1103/PhysRevX.2.041014} {\bibfield  {journal} {\bibinfo
  {journal} {Phys. Rev. X}\ }\textbf {\bibinfo {volume} {2}},\ \bibinfo {pages}
  {041014} (\bibinfo {year} {2012})}\BibitemShut {NoStop}%
\bibitem [{\citenamefont {Thompson}\ \emph {et~al.}(2013)\citenamefont
  {Thompson}, \citenamefont {Tiecke}, \citenamefont {Zibrov}, \citenamefont
  {Vuleti\ifmmode~\acute{c}\else \'{c}\fi{}},\ and\ \citenamefont
  {Lukin}}]{Lukin2013}%
  \BibitemOpen
  \bibfield  {author} {\bibinfo {author} {\bibfnamefont {J.~D.}\ \bibnamefont
  {Thompson}}, \bibinfo {author} {\bibfnamefont {T.~G.}\ \bibnamefont
  {Tiecke}}, \bibinfo {author} {\bibfnamefont {A.~S.}\ \bibnamefont {Zibrov}},
  \bibinfo {author} {\bibfnamefont {V.}~\bibnamefont
  {Vuleti\ifmmode~\acute{c}\else \'{c}\fi{}}}, \ and\ \bibinfo {author}
  {\bibfnamefont {M.~D.}\ \bibnamefont {Lukin}},\ }\href {\doibase
  10.1103/PhysRevLett.110.133001} {\bibfield  {journal} {\bibinfo  {journal}
  {Phys. Rev. Lett.}\ }\textbf {\bibinfo {volume} {110}},\ \bibinfo {pages}
  {133001} (\bibinfo {year} {2013})}\BibitemShut {NoStop}%
\bibitem [{\citenamefont {Sompet}\ \emph {et~al.}(2017)\citenamefont {Sompet},
  \citenamefont {Fung}, \citenamefont {Schwartz}, \citenamefont {Hunter},
  \citenamefont {Phrompao},\ and\ \citenamefont {Andersen}}]{Andersen2017}%
  \BibitemOpen
  \bibfield  {author} {\bibinfo {author} {\bibfnamefont {P.}~\bibnamefont
  {Sompet}}, \bibinfo {author} {\bibfnamefont {Y.~H.}\ \bibnamefont {Fung}},
  \bibinfo {author} {\bibfnamefont {E.}~\bibnamefont {Schwartz}}, \bibinfo
  {author} {\bibfnamefont {M.~D.~J.}\ \bibnamefont {Hunter}}, \bibinfo {author}
  {\bibfnamefont {J.}~\bibnamefont {Phrompao}}, \ and\ \bibinfo {author}
  {\bibfnamefont {M.~F.}\ \bibnamefont {Andersen}},\ }\href {\doibase
  10.1103/PhysRevA.95.031403} {\bibfield  {journal} {\bibinfo  {journal} {Phys.
  Rev. A}\ }\textbf {\bibinfo {volume} {95}},\ \bibinfo {pages} {031403}
  (\bibinfo {year} {2017})}\BibitemShut {NoStop}%
\bibitem [{\citenamefont {Porozova}\ \emph {et~al.}(2019)\citenamefont
  {Porozova}, \citenamefont {Gerasimov}, \citenamefont {Bobrov}, \citenamefont
  {Straupe}, \citenamefont {Kulik},\ and\ \citenamefont
  {Kupriyanov}}]{RSC2019}%
  \BibitemOpen
  \bibfield  {author} {\bibinfo {author} {\bibfnamefont {V.~M.}\ \bibnamefont
  {Porozova}}, \bibinfo {author} {\bibfnamefont {L.~V.}\ \bibnamefont
  {Gerasimov}}, \bibinfo {author} {\bibfnamefont {I.~B.}\ \bibnamefont
  {Bobrov}}, \bibinfo {author} {\bibfnamefont {S.~S.}\ \bibnamefont {Straupe}},
  \bibinfo {author} {\bibfnamefont {S.~P.}\ \bibnamefont {Kulik}}, \ and\
  \bibinfo {author} {\bibfnamefont {D.~V.}\ \bibnamefont {Kupriyanov}},\ }\href
  {\doibase 10.1103/PhysRevA.99.043406} {\bibfield  {journal} {\bibinfo
  {journal} {Phys. Rev. A}\ }\textbf {\bibinfo {volume} {99}},\ \bibinfo
  {pages} {043406} (\bibinfo {year} {2019})}\BibitemShut {NoStop}%
\bibitem [{\citenamefont {Rosenfeld}\ \emph {et~al.}(2011)\citenamefont
  {Rosenfeld}, \citenamefont {Volz}, \citenamefont {Weber},\ and\ \citenamefont
  {Weinfurter}}]{Weinfurter_PRA2011}%
  \BibitemOpen
  \bibfield  {author} {\bibinfo {author} {\bibfnamefont {W.}~\bibnamefont
  {Rosenfeld}}, \bibinfo {author} {\bibfnamefont {J.}~\bibnamefont {Volz}},
  \bibinfo {author} {\bibfnamefont {M.}~\bibnamefont {Weber}}, \ and\ \bibinfo
  {author} {\bibfnamefont {H.}~\bibnamefont {Weinfurter}},\ }\href {\doibase
  10.1103/PhysRevA.84.022343} {\bibfield  {journal} {\bibinfo  {journal} {Phys.
  Rev. A}\ }\textbf {\bibinfo {volume} {84}},\ \bibinfo {pages} {022343}
  (\bibinfo {year} {2011})}\BibitemShut {NoStop}%
\bibitem [{\citenamefont {Berestetskii}\ \emph {et~al.}(1982)\citenamefont
  {Berestetskii}, \citenamefont {Lifshitz},\ and\ \citenamefont
  {Pitaevskii}}]{BERESTETSKII19821}%
  \BibitemOpen
  \bibfield  {author} {\bibinfo {author} {\bibfnamefont {V.}~\bibnamefont
  {Berestetskii}}, \bibinfo {author} {\bibfnamefont {E.}~\bibnamefont
  {Lifshitz}}, \ and\ \bibinfo {author} {\bibfnamefont {L.}~\bibnamefont
  {Pitaevskii}},\ }in\ \href {\doibase
  https://doi.org/10.1016/B978-0-08-050346-2.50006-4} {\emph {\bibinfo
  {booktitle} {Quantum Electrodynamics}}}\ (\bibinfo  {publisher}
  {Butterworth-Heinemann},\ \bibinfo {address} {Oxford},\ \bibinfo {year}
  {1982})\ \bibinfo {edition} {second edition}\ ed.\BibitemShut {Stop}%
\bibitem [{\citenamefont {Sheng}\ \emph
  {et~al.}(2018{\natexlab{b}})\citenamefont {Sheng}, \citenamefont {He},
  \citenamefont {Xu}, \citenamefont {Guo}, \citenamefont {Wang}, \citenamefont
  {Xiong}, \citenamefont {Liu}, \citenamefont {Wang},\ and\ \citenamefont
  {Zhan}}]{Zhan2018}%
  \BibitemOpen
  \bibfield  {author} {\bibinfo {author} {\bibfnamefont {C.}~\bibnamefont
  {Sheng}}, \bibinfo {author} {\bibfnamefont {X.}~\bibnamefont {He}}, \bibinfo
  {author} {\bibfnamefont {P.}~\bibnamefont {Xu}}, \bibinfo {author}
  {\bibfnamefont {R.}~\bibnamefont {Guo}}, \bibinfo {author} {\bibfnamefont
  {K.}~\bibnamefont {Wang}}, \bibinfo {author} {\bibfnamefont {Z.}~\bibnamefont
  {Xiong}}, \bibinfo {author} {\bibfnamefont {M.}~\bibnamefont {Liu}}, \bibinfo
  {author} {\bibfnamefont {J.}~\bibnamefont {Wang}}, \ and\ \bibinfo {author}
  {\bibfnamefont {M.}~\bibnamefont {Zhan}},\ }\href {\doibase
  10.1103/PhysRevLett.121.240501} {\bibfield  {journal} {\bibinfo  {journal}
  {Phys. Rev. Lett.}\ }\textbf {\bibinfo {volume} {121}},\ \bibinfo {pages}
  {240501} (\bibinfo {year} {2018}{\natexlab{b}})}\BibitemShut {NoStop}%
\bibitem [{\citenamefont {Dalibard}\ and\ \citenamefont
  {Cohen-Tannoudji}(1985)}]{Dalibard1985}%
  \BibitemOpen
  \bibfield  {author} {\bibinfo {author} {\bibfnamefont {J.}~\bibnamefont
  {Dalibard}}\ and\ \bibinfo {author} {\bibfnamefont {C.}~\bibnamefont
  {Cohen-Tannoudji}},\ }\href {\doibase 10.1364/JOSAB.2.001707} {\bibfield
  {journal} {\bibinfo  {journal} {J. Opt. Soc. Am. B}\ }\textbf {\bibinfo
  {volume} {2}},\ \bibinfo {pages} {1707} (\bibinfo {year} {1985})}\BibitemShut
  {NoStop}%
\bibitem [{\citenamefont {Happer}(1972)}]{Happer1972}%
  \BibitemOpen
  \bibfield  {author} {\bibinfo {author} {\bibfnamefont {W.}~\bibnamefont
  {Happer}},\ }\href {\doibase 10.1103/RevModPhys.44.169} {\bibfield  {journal}
  {\bibinfo  {journal} {Rev. Mod. Phys.}\ }\textbf {\bibinfo {volume} {44}},\
  \bibinfo {pages} {169} (\bibinfo {year} {1972})}\BibitemShut {NoStop}%
\bibitem [{\citenamefont {Kupriyanov}\ \emph {et~al.}(2005)\citenamefont
  {Kupriyanov}, \citenamefont {Mishina}, \citenamefont {Sokolov}, \citenamefont
  {Julsgaard},\ and\ \citenamefont {Polzik}}]{Mishina2005}%
  \BibitemOpen
  \bibfield  {author} {\bibinfo {author} {\bibfnamefont {D.~V.}\ \bibnamefont
  {Kupriyanov}}, \bibinfo {author} {\bibfnamefont {O.~S.}\ \bibnamefont
  {Mishina}}, \bibinfo {author} {\bibfnamefont {I.~M.}\ \bibnamefont
  {Sokolov}}, \bibinfo {author} {\bibfnamefont {B.}~\bibnamefont {Julsgaard}},
  \ and\ \bibinfo {author} {\bibfnamefont {E.~S.}\ \bibnamefont {Polzik}},\
  }\href {\doibase 10.1103/PhysRevA.71.032348} {\bibfield  {journal} {\bibinfo
  {journal} {Phys. Rev. A}\ }\textbf {\bibinfo {volume} {71}},\ \bibinfo
  {pages} {032348} (\bibinfo {year} {2005})}\BibitemShut {NoStop}%
\bibitem [{\citenamefont {Savard}\ \emph
  {et~al.}(1997{\natexlab{b}})\citenamefont {Savard}, \citenamefont {O'Hara},\
  and\ \citenamefont {Thomas}}]{Savard1997}%
  \BibitemOpen
  \bibfield  {author} {\bibinfo {author} {\bibfnamefont {T.~A.}\ \bibnamefont
  {Savard}}, \bibinfo {author} {\bibfnamefont {K.~M.}\ \bibnamefont {O'Hara}},
  \ and\ \bibinfo {author} {\bibfnamefont {J.~E.}\ \bibnamefont {Thomas}},\
  }\href {\doibase 10.1103/PhysRevA.56.R1095} {\bibfield  {journal} {\bibinfo
  {journal} {Phys. Rev. A}\ }\textbf {\bibinfo {volume} {56}},\ \bibinfo
  {pages} {R1095} (\bibinfo {year} {1997}{\natexlab{b}})}\BibitemShut {NoStop}%
\bibitem [{\citenamefont {Samoylenko}\ \emph {et~al.}(2020)\citenamefont
  {Samoylenko}, \citenamefont {Lisitsin}, \citenamefont {Schepanovich},
  \citenamefont {Bobrov}, \citenamefont {Straupe},\ and\ \citenamefont
  {Kulik}}]{Samoylenko_LPL2020}%
  \BibitemOpen
  \bibfield  {author} {\bibinfo {author} {\bibfnamefont {S.}~\bibnamefont
  {Samoylenko}}, \bibinfo {author} {\bibfnamefont {A.}~\bibnamefont
  {Lisitsin}}, \bibinfo {author} {\bibfnamefont {D.}~\bibnamefont
  {Schepanovich}}, \bibinfo {author} {\bibfnamefont {I.}~\bibnamefont
  {Bobrov}}, \bibinfo {author} {\bibfnamefont {S.}~\bibnamefont {Straupe}}, \
  and\ \bibinfo {author} {\bibfnamefont {S.}~\bibnamefont {Kulik}},\
  }\href@noop {} {\bibfield  {journal} {\bibinfo  {journal} {Laser Physics
  Letters}\ }\textbf {\bibinfo {volume} {17}},\ \bibinfo {pages} {025203}
  (\bibinfo {year} {2020})}\BibitemShut {NoStop}%
\bibitem [{\citenamefont {Schlosser}\ \emph {et~al.}(2001)\citenamefont
  {Schlosser}, \citenamefont {Reymond}, \citenamefont {Protsenko},\ and\
  \citenamefont {Grangier}}]{Grangier_Nature2001}%
  \BibitemOpen
  \bibfield  {author} {\bibinfo {author} {\bibfnamefont {N.}~\bibnamefont
  {Schlosser}}, \bibinfo {author} {\bibfnamefont {G.}~\bibnamefont {Reymond}},
  \bibinfo {author} {\bibfnamefont {I.}~\bibnamefont {Protsenko}}, \ and\
  \bibinfo {author} {\bibfnamefont {P.}~\bibnamefont {Grangier}},\ }\href@noop
  {} {\bibfield  {journal} {\bibinfo  {journal} {Nature}\ }\textbf {\bibinfo
  {volume} {411}},\ \bibinfo {pages} {1024} (\bibinfo {year}
  {2001})}\BibitemShut {NoStop}%
\bibitem [{\citenamefont {Tuchendler}\ \emph {et~al.}(2008)\citenamefont
  {Tuchendler}, \citenamefont {Lance}, \citenamefont {Browaeys}, \citenamefont
  {Sortais},\ and\ \citenamefont {Grangier}}]{Grangier_PRA2008}%
  \BibitemOpen
  \bibfield  {author} {\bibinfo {author} {\bibfnamefont {C.}~\bibnamefont
  {Tuchendler}}, \bibinfo {author} {\bibfnamefont {A.~M.}\ \bibnamefont
  {Lance}}, \bibinfo {author} {\bibfnamefont {A.}~\bibnamefont {Browaeys}},
  \bibinfo {author} {\bibfnamefont {Y.~R.~P.}\ \bibnamefont {Sortais}}, \ and\
  \bibinfo {author} {\bibfnamefont {P.}~\bibnamefont {Grangier}},\ }\href
  {\doibase 10.1103/PhysRevA.78.033425} {\bibfield  {journal} {\bibinfo
  {journal} {Phys. Rev. A}\ }\textbf {\bibinfo {volume} {78}},\ \bibinfo
  {pages} {033425} (\bibinfo {year} {2008})}\BibitemShut {NoStop}%
\bibitem [{\citenamefont {Frese}\ \emph {et~al.}(2000)\citenamefont {Frese},
  \citenamefont {Ueberholz}, \citenamefont {Kuhr}, \citenamefont {Alt},
  \citenamefont {Schrader}, \citenamefont {Gomer},\ and\ \citenamefont
  {Meschede}}]{Meshede2000}%
  \BibitemOpen
  \bibfield  {author} {\bibinfo {author} {\bibfnamefont {D.}~\bibnamefont
  {Frese}}, \bibinfo {author} {\bibfnamefont {B.}~\bibnamefont {Ueberholz}},
  \bibinfo {author} {\bibfnamefont {S.}~\bibnamefont {Kuhr}}, \bibinfo {author}
  {\bibfnamefont {W.}~\bibnamefont {Alt}}, \bibinfo {author} {\bibfnamefont
  {D.}~\bibnamefont {Schrader}}, \bibinfo {author} {\bibfnamefont
  {V.}~\bibnamefont {Gomer}}, \ and\ \bibinfo {author} {\bibfnamefont
  {D.}~\bibnamefont {Meschede}},\ }\href {\doibase 10.1103/PhysRevLett.85.3777}
  {\bibfield  {journal} {\bibinfo  {journal} {Phys. Rev. Lett.}\ }\textbf
  {\bibinfo {volume} {85}},\ \bibinfo {pages} {3777} (\bibinfo {year}
  {2000})}\BibitemShut {NoStop}%
\end{thebibliography}%

\end{document}